# Write Cycling Endurance Exceeding $10^{10}$ in Sub-50 nm Ferroelectric AlScN

Hyunmin Cho[1]†, Yubo Wang[1], Chloe Leblanc[1], Yinuo Zhang[1], Yunfei He[1], Zirun Han[1], Roy H. Olsson III[1]*, Deep Jariwala[1]*

[1]Department of Electrical and Systems Engineering, University of Pennsylvania, Philadelphia, Pennsylvania 19104, USA

*Corresponding author: Deep Jariwala (email: dmj@seas.upenn.edu), Roy H. Olsson III (email: rolsson@seas.upenn.edu)

## Abstract

Wurtzite ferroelectrics, particularly aluminum scandium nitride (AlScN), have emerged as a promising materials platform for non-volatile memories, offering high polarization values exceeding 100 μC/cm². However, their high coercive fields (>3 MV/cm) have limited cycling endurance to ~$10^7$ cycles in previous reports. Here, we demonstrate unprecedented control of polarization switching in AlScN, achieving write cycling endurance exceeding $10^{10}$ cycles—a thousand-fold improvement over previous wurtzite ferroelectric benchmarks. Through precise voltage modulation in 45 nm-thick $Al_{0.64}Sc_{0.36}N$ capacitors, we show that while complete polarization reversal ($2P_r \approx 200$ μC/cm²) sustains ~$10^8$ cycles, partial switching extends endurance beyond $10^{10}$ cycles while maintaining a substantial polarization (>30 μC/cm² for $2P_r$). This exceptional endurance, combined with breakdown fields approaching 10 MV/cm in optimized 10 μm diameter devices, represents the highest reported values for any wurtzite ferroelectric. Our findings establish a new paradigm for reliability in nitride ferroelectrics, demonstrating that controlled partial polarization and size scaling enables both high endurance and energy-efficient operation.

# Introduction

The exponential growth in data processing demands has intensified the search for energy-efficient, high-endurance, non-volatile memories. While conventional memory technologies face fundamental scaling and endurance limitations[1,2], ferroelectric materials offer a promising solution through their non-volatile polarization switching and CMOS compatibility[3,4]. However, achieving both high endurance (>$10^{10}$ cycles) and reliable operation remains a critical challenge for the practical implementation of non-volatile memory technologies[5].

The discovery of ferroelectricity in wurtzite-structured aluminum scandium nitride (AlScN) in 2019 introduced a fundamentally new class of polar materials[6]. Unlike conventional oxide ferroelectrics which rely on oxygen vacancy ordering ($HfO_2$ based ferroelectrics) or B-site cation displacement (PZT), wurtzite ferroelectrics achieve polarization through intrinsic ionic charge separation in a non-centrosymmetric lattice[7,8]. This distinct mechanism enables remarkably high remnant polarization exceeding 100 μC/cm² in AlScN—significantly higher than the typical range of 10-40 μC/cm² observed in $HfO_2$-based ferroelectrics[3,8]. Moreover, AlScN can be deposited at back-end-of-line (BEOL)-compatible temperatures below 400°C while maintaining reliable ferroelectric switching in films as thin as 5 nm[9]. These properties, combined with demonstrated operation at temperatures up to 600°C[10], position wurtzite ferroelectrics as promising candidates for next-generation non-volatile memory applications.

The implementation of wurtzite ferroelectrics in practical memory devices faces two key challenges. First, previously reported AlScN devices show limited endurance, typically failing before $10^7$ switching cycles[11,12]. Second, the high coercive fields necessary for complete polarization switching (>3 MV/cm) lead to significant power consumption and accelerated device degradation[7]. These limitations are particularly pronounced in scaled devices, where local defects can dominate switching behavior and lead to premature breakdown[13].

In this work, we demonstrate a breakthrough in AlScN ferroelectric device reliability through controlled partial polarization switching and systematic size scaling. Using precisely modulated voltage pulses in 45 nm-thick $Al_{0.64}Sc_{0.36}N$ capacitors, we achieve write cycling endurance exceeding $10^{10}$ cycles while maintaining substantial remnant polarization above 30 μC/cm². This represents a thousand-fold improvement over previous benchmarks for wurtzite ferroelectrics[11,12]. By reducing device dimensions to 10 μm diameter, we achieve breakdown fields approaching 10 MV/cm, establishing new performance standards for nitride ferroelectrics. Our findings reveal that controlled partial polarization, combined with optimal device scaling, provides a pathway to simultaneously achieve high endurance and energy efficiency in ferroelectric memories.

# Main

The schematic in **Figure 1a** illustrates an $Al_{0.64}Sc_{0.36}N$ capacitor specifically designed to minimize stress on the $Al_{0.64}Sc_{0.36}N$ layer during measurements. This reduction in stress was achieved to avoid the pressure of the probe tip with the via contact. The device features an Al (50 nm) bottom electrode, an $Al_{0.64}Sc_{0.36}N$ (45 nm) layer, and an Al (50 nm) top electrode, all deposited in-situ to ensure minimal interface defects. Additionally, Pd/Ti/Cr (250 nm) was deposited as the via contact material, and $SiO_2$ (200 nm) was utilized below the contact pad to sustain the capacitor structure. This architecture isolates the top and bottom contact pads with a via structure, effectively reducing unwanted capacitance and leakage currents. During operation, voltage was applied to the bottom electrode, while the top electrode was grounded, allowing stable current measurements from the top contact. Furthermore, this design minimizes current detours or breakdown paths between the top and bottom electrode, ensuring reliable operation under consistent measurement conditions.

**Figure 1b** and **1c** shows optical microscope images from the top view. Various top electrode sizes were prepared simultaneously on the same sample to minimize sample fabrication variations. Additionally, the bottom contact pad was positioned consistently for each set of capacitors to ensure reliable measurement results. With this well-prepared device, the current density–voltage (*J-V*) characteristics were measured to evaluate the electrical behavior as shown in **Figure 1d**. The measurements were conducted with a standard bipolar voltage pulse with a linear ramp-up and ramp-down sequence, configured as a 10 kHz triangular pulse. To ensure consistency, capacitors with varying diameters were tested. The results demonstrated high reliability, showing minimal variation regardless of diameter. An increase in leakage current was observed for positive voltages as the applied voltage approached higher values. This behavior led to an overestimation of the $2P_r$, making accurate measurements infeasible for positive applied switching voltages. Therefore, the analysis focused on the data from the negative voltage switching, where leakage effects were minimized[14]. This approach enabled a reliable determination of the device's polarization and switching characteristics. These *J-V* characteristics are similar to our previous reports[4,10]. Additionally, we conducted quasi direct current-voltage (DC-IV) measurements with various diameters, as shown in **Supplementary Figure S1**.

To quantitatively analyze $2P_r$, Positive-Up-Negative-Down (PUND) pulsed measurements were performed at a frequency of 10 kHz, using a square pulse width of 50 μs. This methodology is highly effective for isolating the ferroelectric switching current response by minimizing the impact of leakage currents[15]. Consequently, it provides a precise measurement of the intrinsic ferroelectric properties. In this study, the $2P_r$ extracted using the PUND method is defined as the 'conventional $2P_r$'.

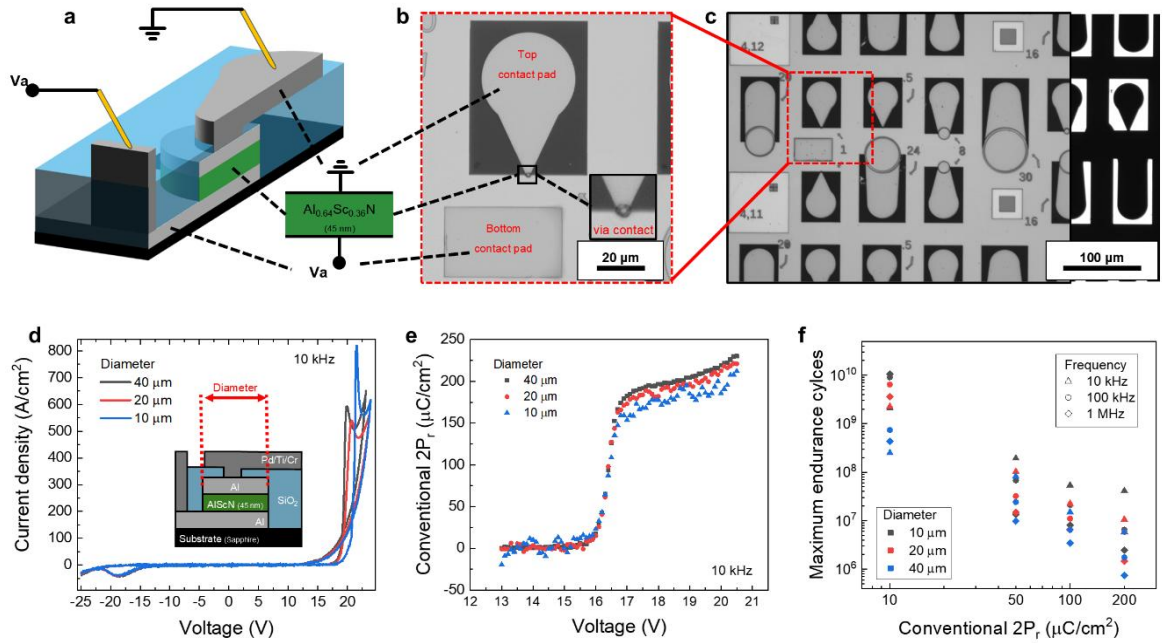

**Figure 1| Schematic of AlScN capacitor and its electrical characteristics. (a)** Schematic illustration of the $Al_{0.64}Sc_{0.36}N$ ferroelectric capacitor with a via-contact structure. **(b, c)** Optical microscope images of the ferroelectric capacitor. The number on the image indicates the radius of top electrode. **(d)** J–V hysteresis loops, **(e)** ferroelectric polarization response to applied voltage pulses, and **(f)** maximum endurance cycles vs. conventional $2P_r$ for various diameters (10, 20, and 40 μm).

**Figure 1e** shows the relationship between conventional $2P_r$ and applied voltage. This test was conducted across capacitors with varying top electrode diameters. Switching of ferroelectric dipoles were initiated at a voltage pulse of 16 V, with conventional $2P_r$ rapidly approaching a saturation value of ~200 μC/cm². Within the voltage range where conventional $2P_r$ transitions from onset to saturation, partial polarization was observed. This effect is due to incomplete domain switching within the applied voltage range[16-19]. Gradual alignment of ferroelectric domains during this process provides key insights into partial polarizations of AlScN. Therefore, this behavior reflects the dynamic switching characteristics of the material. Notably, the partial polarization remained stable across varying top electrode diameters, demonstrating high reproducibility and enabling direct comparison of results independent of electrode size. **Supplementary Figure S2** and **S3** presents the corresponding time-dependent current response. Additionally, **Supplementary Information S1** specifically focuses on PUND results to explain the partial polarization switching, providing further insights into this phenomenon.

Partial polarization has emerged as a promising mechanism for advancing multistate memory and neuromorphic devices[4,20-23]. Despite its recognized potential, achieving stable and controllable intermediate polarization states remains a critical challenge. Notably, AlScN presents an exceptional ability to sustain stable partial polarization due to its distinct ferroelectric mechanism[18]. Building on these characteristics, this study explores an innovative approach to enhance endurance cycles by leveraging the intrinsic ferroelectric behavior of AlScN, thereby pushing the boundaries of device performance and reliability.

To further investigate this behavior, additional measurements were performed on a 40 μm capacitor at a different frequency as shown in **Supplementary Figure S2**. As the frequency increases, an increase in operation voltage was observed. The increase of operation voltage refers to the applied voltage necessary to obtain the same conventional $2P_r$. Additionally, a difference in the slope of conventional $2P_r$ from onset to saturation was detected. This indicates a distinct frequency-dependent modulation of domain switching. The observation of increased coercive voltage ($V_C$) is consistent with reports in previous literature[6,9,13,24,25]. This suggests that the switching kinetics are influenced by the interplay between external field dynamics and the intrinsic material response. This interplay highlights the importance of further exploring frequency-dependent polarization behavior. Understanding these effects could provide deeper insights into domain switching mechanisms.

**Figure 1f** shows the relationship between conventional $2P_r$ and maximum endurance cycles in AlScN capacitors. To comprehensively investigate this relationship, we conducted additional tests by varying both the frequency of the fatigue voltage pulses and the diameter of the capacitors' top electrode. Here, we must pay attention on the x-axis to conventional $2P_r$ values lower than 200 μC/cm², such as 10, 50, and 100 μC/cm². These are the results of partial polarization because of applying a switching voltage lower than $V_C$. However, such partial polarization values cannot be accurately extracted using the conventional PUND method which is only suitable for full polarization measurements. Because, based on the explanation provided in **Supplementary Information S2**, the conventional $2P_r$ values (10, 50, and 100) are always lower than the actual partially switched $2P_r$. However, conventional $2P_r$ values are used on the x-axis for consistent comparison. In addition, we define the actual amount of partially or fully switched polarization as 'intrinsic $2P_r$' in this context.

To test the endurance of partially or fully switched AlScN, each fatigue iteration consists of one positive and one negative pulse with identical time of pulse duration. A detailed pulse train configuration is depicted in **Supplementary Figure S4**. During the test, the voltage pulse amplitudes are carefully adjusted to keep the measured conventional $2P_r$ close to a specific $2P_r$ value. Here, we define the specific $2P_r$ targeted in the program as the 'preset $2P_r$'. Therefore, the x-axis in **Figure 1f** represents both the conventional 2P$_r$ and the preset 2P$_r$. This equivalence is ensured by a well-developed program that maintains consistency between the conventional (measured) and preset (programmed) 2P$_r$ values with minimal deviation.

A pivotal observation from these experiments is that a reduced conventional 2P$_r$ markedly enhances the endurance of the capacitors. This inverse relationship shows that as the conventional 2P$_r$ increases, the maximum endurance cycles diminish accordingly. Remarkably, capacitors subjected to a conventional 2P$_r$ of 10 μC/cm² exhibit endurance exceeding 10 billion cycles, a performance that significantly outpaces prior reports. Furthermore, reducing the top electrode diameter contributes to a noticeable improvement in endurance. Detailed explanations are discussed in **Figure 3**. Further, it is also noteworthy that even for full conventional 2P$_r$ switching equaling ~200 μC/cm², the 10 μm diameter capacitors last up to $10^8$ cycles at a 10 kHz frequency, again surpassing any published report for a thin AlScN film by more than an order of magnitude[11,12]. To recognize the trend easily, the plots were subdivided and prepared in **Supplementary Figure S5.**

Finally, this trend graph also reveals that the relationship between frequency and endurance is influenced by the magnitude of the conventional 2P$_r$, which is maintained close to preset 2P$_r$. Generally, higher frequencies result in increased endurance performance[26,27]. However, our results show different tendencies depending on the conventional 2P$_r$ due to the self-adjusted voltage. For larger conventional 2P$_r$ values, endurance tends to decrease with increasing frequency, indicating an inverse correlation. In contrast, at lower conventional 2P$_r$ values, such as 10 μC/cm², endurance improves as frequency increases, as shown in **Supplementary Figure S6**. Additionally, the general trend of endurance dependence on the frequency can be observed from our test results, as explained in **Figure 2**.

As frequency increases, V$_C$ also increases, as shown in **Supplementary Figure S2** and **Figure S7**. Consequently, regardless of the preset 2P$_r$, achieving a conventional 2P$_r$ close to the preset 2P$_r$ value requires a higher applied voltage at higher frequencies. Such an increase in applied voltage results in two negative effects. Increased applied voltage introduces additional electrical stress, accelerating dielectric breakdown due to a large electric field[28]. Furthermore, it raises peak current density, leading to increased electromigration and Joule heating, accelerating material degradation again[29,30]. These negative effects of frequency reduce endurance, particularly at high preset 2P$_r$, which requires a higher applied voltage and leads to increased current density.

However, high frequency increases not only the applied voltage but also the breakdown voltage (V$_{BD}$). At high frequency, pulse width time decreases. This shortens the duration of electrical stress exposure. As a result, the V$_{BD}$ increases. Moreover, a reduction in the preset 2P$_r$ leads to a relatively lower applied voltage at a given frequency. This results in a higher ratio of V$_{BD}$ to applied voltage, reducing electrical stress and limiting dielectric breakdown. Additionally, this lower applied voltage decreases the peak current density, minimizing electromigration and Joule heating significantly. According to previous reports[29], decreases in current density are possible to induce dramatic suppression of these negative effects. This is because current based stress generally has a positive correlation with the square of the current. Therefore, these mechanisms enhance endurance, as demonstrated in **Supplementary Figure S6b**. The improvement is particularly evident in smaller devices, where reduced defect density further decreases sensitivity to electrical stress. **Figure 3** provides further insights into this effect. These results highlight the importance of optimizing frequency to balance its competing effects on endurance, particularly concerning partial polarization switching.

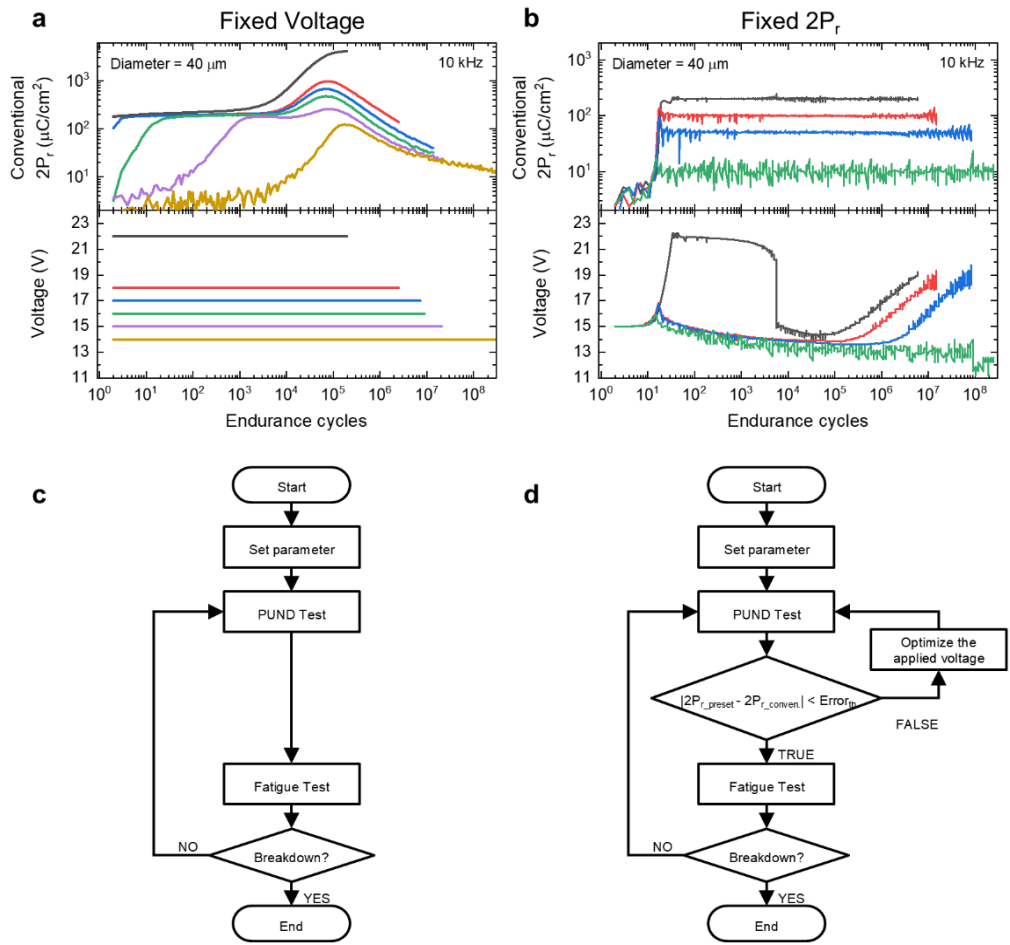

**Figure 2| The results of endurance test.** Endurance cycles under **(a)** constant applied voltage pulses and **(b)** adjusted applied voltage pulses to maintain a constant conventional $2P_r$. **(c, d)** flow charts illustrate the endurance test algorithm designed to **(c)** a fixed voltage and **(d)** a fixed $2P_r$.

**Figure 2** presents a detailed analysis of endurance characteristics under varying voltage conditions. **Figure 2a** provides the results obtained from endurance tests conducted at fixed voltages ($V_{fixed}$). The bottom panel of the figure represents the applied voltage, while the top panel shows the evolution of conventional $2P_r$ throughout the testing cycles. Consistent with prior studies[11,31-35], the wake-up effect was observed in our AlScN capacitors, characterized by an increase in conventional $2P_r$ after repeated applied voltage cycling. This phenomenon is followed by progressive fatigue leading to device failure. Most conventional endurance tests typically apply voltages exceeding $V_C$. Therefore, most results show a reduction in conventional $2P_r$, which continues to decline over the last one to three decades of the endurance test. This degradation raises concerns about the reliability of ferroelectricity and suggests that endurance values may be overestimated. However, our study employs a wide range of $V_{fixed}$ for partial and full switching. This approach explores the potential of partial polarization in enhancing device endurance and mitigating degradation. The black curve in **Figure 2a** corresponds to a voltage which is much higher than $V_C$, exhibiting behavior consistent with high-performance ferroelectric devices reported in previous literature[26,31]. This includes a short wake-up phenomenon and a gradual degradation, similar to previously observed trends[11,31-35]. These consistent results reinforce the reliability of our endurance measurements and the high quality of our AlScN. Moreover, further results of endurance cycles under various frequencies are shown in **Supplementary Figure S8**. Additionally, endurance at 22 V across different frequencies (shown in **Figure 2a** and **Supplementary Figure S8)**. shows explicitly the same positive correlation observed in previous reports[26,27].

To investigate the influence of partial switching on endurance, progressively lower voltages were applied, as indicated by the colored curves in **Figure 2a**. The simple flowchart used for testing the $V_{fixed}$ endurance test is described in **Figure 2c**. These tests were performed with a constant applied voltage until breakdown. A clear trend emerges where lower voltage correlates with enhanced endurance, highlighting the potential reliability benefits of partial polarization. For example, the purple curve exhibits minimal initial conventional $2P_r$, indicating suppressed polarization switching at the early stages due to the small amplitude of the applied voltage pulse. This endurance behavior can be divided into four distinct phases. This progression of conventional $2P_r$ evolution is important and requires further discussion. For a comprehensive analysis, refer to **Supplementary Figure S9 and Information S3**. However, a brief description of each phase follows. The first phase of the $V_{fixed}$ test begins with the onset of polarization switching, characterized by a gradual increase in conventional $2P_r$. The first phase of the $V_{fixed}$ test is the result of the partial wake-up phenomenon. This is followed by second phase of the $V_{fixed}$ test, where conventional $2P_r$ remains constant at a stable state. Subsequently, the third phase of the $V_{fixed}$ test is marked by a pronounced increase in conventional $2P_r$, which is progressively overestimated due to increasing leakage current. Finally, the fourth phase of the $V_{fixed}$ test is characterized by a reduction in conventional $2P_r$ due to degradation of ferroelectricity. The fourth phase of the $V_{fixed}$ test is consistent with the fatigue phenomenon reported in previous studies[11,31-35]. Even under sub-coercive voltages, the wake-up effect remains evident[36]. The presence of a partial wake-up phase (first phase of the $V_{fixed}$ test) delicately activates the device without causing unnecessary high electrical stress. This controlled activation contributes to prolonged endurance performance. Degradation of AlScN begins when increased leakage is initiated. However, it progresses at a slower rate compared to devices subjected to higher voltages. These observations suggest that fine-tuning the applied voltage can significantly enhance the endurance and operational reliability of ferroelectric devices.

In **Figure 2b**, we maintained consistent conventional $2P_r$ values during endurance tests by carefully modulating the applied voltage. This approach provides several advantages. First, it introduces a novel method for controlling polarization in ferroelectric materials, enabling broader utilization of AlScN. Fine-tuning the applied voltage enhances adaptability for industrial applications. This approach makes AlScN more suitable for applications demanding different $2P_r$ values from a single material. Second, this method ensures high reliability by stabilizing $2P_r$ values without fluctuations throughout operation. This stability helps engineers design circuits with a clear understanding of the behavior of AlScN. This approach prevents performance degradation and extends device lifespan. Finally, we achieved significant improvements in endurance cycles. Our approach maintains stable conventional $2P_r$ throughout the test duration, demonstrating consistent performance over extended cycles.

The applied voltage was carefully controlled using an algorithm shown in **Figure 2d**, which illustrates the programming flowchart implemented with the Keithley Kult software. To keep the conventional $2P_r$ response stable, the evolution of voltage is divided into four phases. More details can be found in **Supplementary Figure S9 and Information S3**. During the initial phase of the Fixed $2P_r$ ($2P_{r\_Fixed}$) test, the starting voltage cannot be predetermined. To raise the conventional $2P_r$ response approach the preset $2P_r$, the applied voltage was carefully increased, while the algorithm ensured a rapid rise within tens of cycles to minimize stress. This was followed by the second and third phases of the $2P_{r\_Fixed}$ test, which exhibit distinct characteristics depending on the preset $2P_r$ value. For instance, in the case of the black curve in **Figure 2b**, the second phase of the $2P_{r\_Fixed}$ test shows a rough decrease in voltage following an initial increase. Subsequently, in the third phase, a sharp voltage drop is observed, followed by a continued gradual decline. These trends align with the behaviors observed in the second (stable) and third (leakage) phases of the $V_{fixed}$ test (detailed in **Supplementary Figure S9 and Information S3**). In contrast, for lower preset $2P_r$ values, the sharp voltage drop is not apparent. Instead, the second and third phases of the $2P_{r\_Fixed}$ test progress simultaneously at a slower rate due to the partial wake-up process of the ferroelectric system. During this process, both partial wake-up and stabilization emerges together. The gradual rise in adjusted applied voltage in AlScN contributes to enhanced endurance by mitigating stress-induced degradation and ensuring long-term operational reliability. Notably, as preset $2P_r$ decreases, the distinction between the partial wake-up and stabilization phases becomes increasingly indistinct, as illustrated in **Supplementary Figure S9c**. As the device approaches the end of its operational lifespan, the applied voltage is incrementally increased to sustain the conventional $2P_r$ response, a period corresponding to the fourth phase of $2P_{r\_Fixed}$ (fatigue phase). The feedback loop is designed to keep the conventional $2P_r$ response as close as possible to preset $2P_r$, as shown in **Figure 2d**. Adjustment sensitivity parameters can be controlled by setting the error threshold ($Error_{th}$) parameter and tuning various parameters into the optimization process. These parameters are configured before the test begins. Moreover, the pulse configuration parameters for the PUND and fatigue pulses are configured together. (Detailed description in **Supplementary Figure S4**)

**Figure 3** exhibits the dependence of ferroelectric behavior in AlScN capacitors on the top electrode diameter. In **Figure 3a**, the relationship between electrode diameter and two critical parameters is shown. The parameters under investigation are the coercive ($E_C$) and breakdown electric fields ($E_{BD}$). These critical values are divided by the 45 nm thickness of the AlScN layer, corresponding to the $V_C$ and $V_{BD}$. $V_C$ values were extracted from PUND measurements (refer to **Supplementary Figure S10**). The $E_C$ values exhibit negligible variation across different electrode diameters, suggesting stable polarization switching characteristics. On the other hand, $E_{BD}$ exhibits a significant increase as the electrode diameter decreases[9]. Smaller electrode dimensions lead to a higher $E_{BD}/E_C$ ratio, which is positively linked to improved endurance in ferroelectric capacitors. The increase in $E_{BD}$ with smaller electrode sizes is primarily due to a lower chance of defects forming conductive paths between the top and bottom electrodes. While the intrinsic defect density within the capacitor remains constant, the total number of defects capable of initiating breakdown diminishes with a reduction in electrode area[37]. This decrease in potential failure sites contributes to improved device reliability and extended operational endurance[38]. Our observations suggest that reducing the electrode diameter allows for better defect control[37], which in turn directly influences the breakdown performance[38]. This confirms that fewer defects improve breakdown behavior, reinforcing that smaller electrodes enhance endurance. Moreover, we repeated the measurements at different voltage frequencies (refer to **Supplementary Figure S2 and S7**). As reported in previous research[6,9,13,24,25], the $E_C$ increased with frequency. Our measurements also show that $E_{BD}$ increases with frequency, consistent with a previous report[14].

Furthermore, **Supplementary Figure S11** presents a detailed analysis of the relationship between conventional $2P_r$ and $E_C$ or adjusted applied voltage, with variations in electrode diameter and frequency. This comparison highlights key trends in switching characteristics, offering insights into the effects of scaling and frequency dependence in AlScN capacitors.

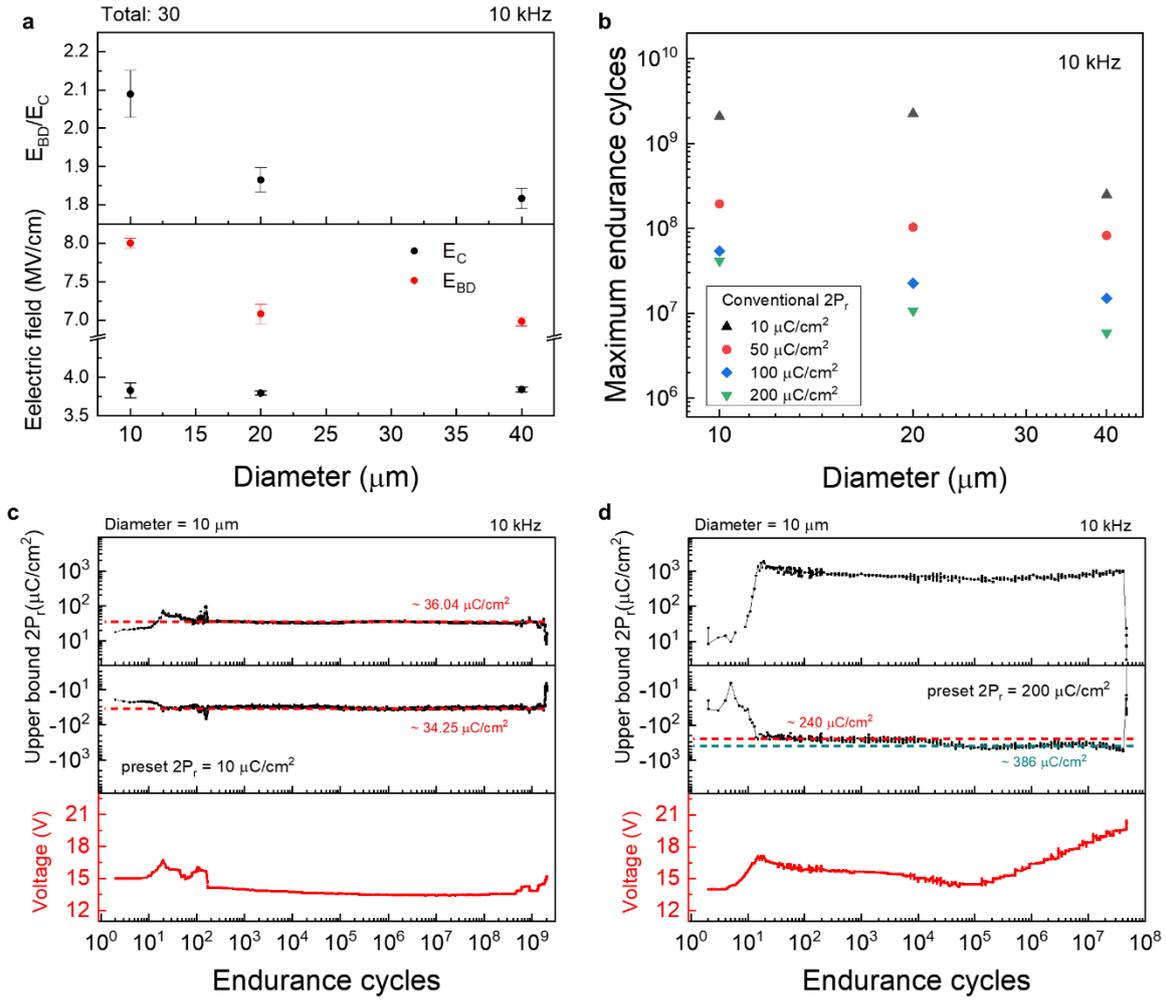

**Figure 3 | Dependence of Ferroelectric Behavior on Electrode Diameter in AlScN Capacitors. (a)** shows the relationship between electrode diameter and ferroelectric characteristics such as $E_C$, $E_{BD}$ (bottom panel), and their ratio (top panel). **(b)** displays the trend of endurance cycles as a function of electrode diameter. **(c, d)** represent endurance test results corresponding to preset $2P_r$ values of 10 μC/cm² and 200 μC/cm², respectively, with their upper bound $2P_r$ values of 34 μC/cm² and 200 μC/cm².

In **Figure 3b**, the data from **Figure 1f** is reorganized to focus on the relationship between electrode diameter and maximum endurance cycles. As expected, the maximum endurance cycles increased as the diameter decreased, which corresponds with the trends observed in **Figure 1f**. This suggests that there is room for additional improvements in endurance. To further investigate this behavior, additional reorganized plots were prepared in **Supplementary Figure S12**. The endurance enhancement with reduced diameter follows the trend observed across all frequency conditions.

Detailed endurance outcomes for 10 μm diameter capacitors, representing both the highest (preset $2P_r$ = 200 μC/cm²) and lowest (preset $2P_r$ = 10 μC/cm²) maintained conventional $2P_r$ values, are presented in **Figures 3c** and **3d**, respectively. These figures illustrate two representative examples of 10 μm diameter capacitors that exhibit different conventional $2P_r$ values. Notably, the capacitor maintaining a conventional $2P_r$ of 10 μC/cm² at 10 kHz demonstrates endurance exceeding 1 billion cycles, underscoring the exceptional performance of AlScN capacitors. For plots of the PUND current response under different conditions measured at intermediate points during endurance tests, please refer to **Supplementary Figure S13** and **S14**.

**Figures 3c** and **3d** display integrated current density or charge density (P & N) instead of conventional $2P_r$ on the y-axis. Conventional PUND methods, such as subtracting U from P or D from N, underestimate the $2P_r$ at low voltages due to incomplete domain switching. This is because some dipoles must still be partially switched during the U and D pulses (the second pulse of each polarity). We define P and N, divided by area (P/A and N/A) as the 'upper bound $2P_r$'. This method may slightly overestimate $2P_r$ due to leakage currents or RC delays. However, these effects are minimal at low applied voltages. Therefore, in **Figure 3c**, the upper bound $2P_r$ of approximately 34.25 μC/cm² ensures high measurement fidelity. On the other hand, the fully switched intrinsic $2P_r$ requires relatively higher voltages, increasing the risk for overestimating the switched polarization. However, before the AlScN fully enters the leakage or degradation phases, this overestimation is much lower. This is because the relatively low voltage applied throughout the entire test is enough to obtain the conventional $2P_r$, reaching the preset $2P_r$. Before the leakage and degradation phase, the upper bound $2P_r$ (~240 μC/cm²) approaches the expected value (200 μC/cm² of conventional $2P_r$), indicating that leakage and other overestimation factors are minimized. This observation confirms the excellent quality of our AlScN. This finding shows that our AlScN capacitor exhibits minimal leakage and few defects due to optimized sputtering and carefully designed capacitors. This reflects excellent stability and outstanding material quality. Further details and analyses are in **Supplementary Information S2**. Additionally, **Supplementary Figure S15** presents our best endurance test results. This plot highlights the remarkable durability of our device, further validating the robustness of our findings.

We compare the endurance performance of our AlScN with previously reported ferroelectric materials. Here, we use the conventional $2P_r$ for consistency in comparison with previous reports. As shown in **Figure 4a**, our results cover a wide range of $2P_r$ values, demonstrating a unique capability that has not been previously reported. Unlike prior studies, which did not demonstrate controllable $2P_r$, our findings allow comparisons not only within similar $2P_r$ regimes but also with other ferroelectric materials such as PZT and $HfO_2$-based ferroelectric materials. In addition, the intrinsically high $2P_r$ level of AlScN enables us to explore various $2P_r$ states. A lower intrinsic maximum $2P_r$ level would severely limit the ability to access multiple $2P_r$ states. Our results also confirm that our endurance properties surpass those of other AlN-based ferroelectrics. Although the endurance performance at low $2P_r$ remains below the best-reported values, it remains competitive. Moreover, as mentioned above, further performance enhancements are possible. AlScN is emerging as a stronger candidate than traditional ferroelectric materials. The y-axis of effective $2P_r$ represents the most dominant $2P_r$ value observed throughout the endurance test, consistent with previous reports[55].

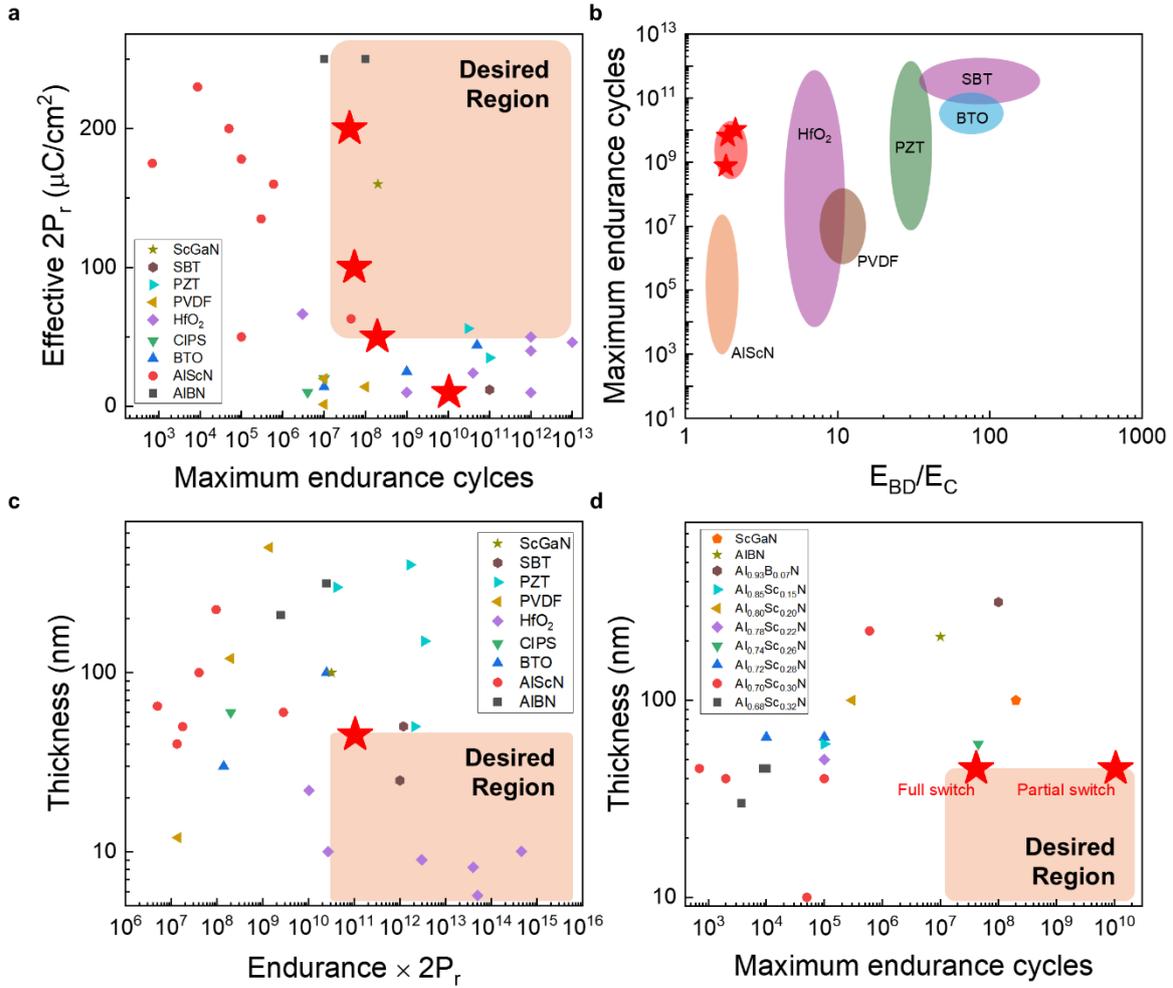

**Figure 4 | Benchmarking of ferroelectric endurance performance.** A systematic comparison of AlScN endurance characteristics with previously reported ferroelectric materials. The red markers represent data from this study. Among the multiple data points collected under identical x-axis conditions, the most representative dataset was selected. Benchmarking maximum endurance cycles against effective **(a)** $2P_r$ and **(b)** $E_{BD}/E_C$ ratio. **(c)** Benchmark of normalized endurance, obtained by multiplying endurance by $2P_r$ for a reasonable comparison against thickness across different ferroelectric materials. **(d)** Thickness-dependent endurance benchmarking among Nitride-based ferroelectrics. Data from: AlScN ($Al_{1-x}Sc_xN$[4,11,14,32,33,39–46]), AlBN ($Al_{1-x}B_xN$[36,47]), ScGaN ($Sc_xGa_{1-x}N$[48]), $HfO_2$[49–56], PZT ($Pb(Zr_xTi_{1-x})O_3$[57–60]), PVDF (poly(vinylidene fluoride))[61–63], BTO ($BaTiO_3$[64–67]), CIPS ($CuInP_2S_6$[68,69]), SBT ($SrBi_2Ta_2O_9$[70, 71]).

As illustrated in **Figure 4b**, our data follows the trend showing that increased $E_{BD}/E_C$ ratio corresponds to increased endurance. In this comparison plot we can find a significant enhancement in both parameters. The $E_{BD}/E_C$ ratio is higher than in previous reports about AlScN, indicating superior material quality. Moreover, endurance improvement is nearly 2 - 3 decades greater than previously reported values of AlScN. In **Figure 4c**, the y-axis is normalized to provide a clearer evaluation metric. Normalized endurance is an important factor in evaluating ferroelectric materials, as endurance and polarization are often inversely correlated. High polarization accelerates fatigue due to excessive domain wall motion and defect activation, whereas low polarization extends endurance while limiting overall ferroelectric performance. To balance this trade-off, multiplying endurance by $2P_r$ provides a more reasonable comparison metric. Despite these challenges, our results demonstrate significantly superior performance compared to other AlScN studies. For a more detailed comparison, in **Figure 4d**, each data point for AlN based ferroelectric materials is scattered in a single plot. Our study presents substantial advantages in terms of both thickness and endurance. Even for fully switched $2P_r$ endurance results, our results establish new benchmarks. Additionally, when partial polarization is considered, significant improvements of approximately three orders of magnitude higher than previous records are observed. These findings position AlScN as a transformative material in ferroelectric endurance research.

## Conclusion

Our work establishes a new paradigm for achieving ultra-high endurance in wurtzite nitrides ferroelectric devices through controlled partial polarization switching. By demonstrating write cycling endurance exceeding $10^{10}$ cycles in AlScN—a thousand-fold improvement over previous benchmarks—we overcome a fundamental limitation in wurtzite ferroelectric reliability while maintaining switched $2P_r > 30$ μC/cm². The combination of partial polarization control and device scaling not only extends endurance but also enables operation at reduced voltages, addressing both reliability and energy efficiency challenges. These findings reveal that wurtzite ferroelectrics can surpass the endurance-polarization trade-off traditionally seen in oxide ferroelectrics, opening new opportunities for practical nonvolatile memories. Beyond memory applications, our approach of controlled partial switching provides a general strategy for enhancing reliability in other emerging ferroelectric materials and devices. This work transforms our understanding of polarization dynamics in wurtzite structures while establishing engineering principles for next-generation ferroelectric technologies.

# Methods

## Substrate Preparation and $Al_{0.64}Sc_{0.36}N$ Deposition

The fabrication process began with the deposition of an $Al_{0.64}Sc_{0.36}N$ thin film on a 6-inch sapphire wafer with a C-plane (0001) orientation off M-plane (1-100) by 0.2 ± 0.1 degrees. A 50 nm Al layer was first deposited at 150°C. A 45 nm AlScN layer was then co-sputtered at 350°C with a nitrogen flow of 30 sccm, utilizing 900 W and 700 W power for 100 mm diameter Al and Sc, respectively. Finally, a 50 nm aluminum layer was deposited at 150°C as the top capping layer. The entire deposition process was conducted in situ to prevent oxidation and maintain film integrity using an Evatec Clusterline 200 II system. This controlled deposition ensured a high-quality crystalline structure with minimal defects. The composition of $Al_{0.64}Sc_{0.36}N$ was carefully optimized to balance $P_r$ and $E_C$, ensuring robust ferroelectric behavior.

## Device Fabrication

The first electron beam lithography (EBL) process was performed to define a pattern facilitating via contact formation while preserving the top Al layer using an EBPG5200+, Raith. A 50 nm thick chromium (Cr) layer was deposited using an E-Beam sputtering system (PVD 75, Kurt J. Lesker) to protect the top Al electrode. A second EBL step was employed to pattern the top electrode structure, followed by inductively coupled plasma (ICP) etching to selectively remove the top Al layer using a Cobra PlasmaPro 100, Oxford Instruments. This Al layer, originally serving as a capping layer, remained functional as the top electrode, minimizing interfacial defects between AlScN and the top Al electrode. The AlScN film and the bottom electrode were subsequently etched using the same EBL and ICP etching process. A 200 nm thick $SiO_2$ layer was deposited using plasma-enhanced chemical vapor deposition (PECVD) to provide insulation using a PlasmaLab 100, Oxford. Following this, via contacts and contact pads were patterned by EBL process. The etching process was conducted by Reactive-ion etching (RIE) process using an 80 Plus, Oxford Instrument. Finally, to fill the via hole and define the contact pad, Pd/Ti were deposited using a E-beam sputtering process.

## Electrical characterization measurement

The electrical characterization was carried out at room temperature using a probe station interfaced with a Keithley 4200A-SCS semiconductor parameter analyzer (Tektronix Inc.).

**Acknowledgements**

The authors acknowledge support from the Intel SRS program. D.J. also acknowledges partial support from the Office of Naval Research (ONR) Nanoscale Computing and Devices program (N00014-24-1-2131) and the Air Force Office of Scientific Research (AFOSR) GHz-THz program grant number FA9550-23-1-0391. A portion of the sample fabrication, assembly, and characterization were carried out at the Singh Center for Nanotechnology at the University of Pennsylvania, which is supported by the National Science Foundation (NSF) National Nanotechnology Coordinated Infrastructure Program grant NNCI-1542153.

**Author contributions**

D.J. and R.H.O. conceived the idea and designed the overall experiments. H.C. developed the code for the endurance cycle test. H.C., Y.W., Y.H., and Z.H. conducted the current-voltage measurements. R.H.O. supervised the AlScN growth process. H.C., C.L., and Y.Z. deposited the AlScN. H.C. designed and carried out the device fabrication processes. D.J., R.H.O., and H.C. analyzed the data, prepared the figures, and wrote the manuscript. All authors contributed to the discussion, analysis of the results, and manuscript writing.

**Competing interests**

The authors declare no competing interests.

# References

1. T. N. Theis, H.-S. P. Wong, The End of Moore's Law: A New Beginning for Information Technology, *Computing in Science & Engineering*, **19**, 41-50, (2017)
2. S. Salahuddin, K. Ni, S. Datta, The era of hyper-scaling in electronics., *Nat Electron*, **1**, 442–450, (2018)
3. T. Mikolajick, S. Slesazeck, H. Mulaosmanovic, M. H. Park, S. Fichtner, P. D. Lomenzo, M. Hoffmann, U. Schroeder, Next generation ferroelectric materials for semiconductor process integration and their applications., *J. Appl. Phys.*, **129**, (2021)
4. KH Kim, S. Oh, M. M. A. Fiagbenu, J. Zheng, P. Musavigharavi, P. Kumar, N. Trainor, A. Aljarb, Y. Wan, H. M. Kim, K. Katti, S. Song, G. Kim, Z. Tang, JH Fu, M. Hakami, V. Tung, J. M. Redwing, E. A. Stach, R. H. Olsson III, D. Jariwala, Scalable CMOS back-end-of-line-compatible AlScN/two-dimensional channel ferroelectric field-effect transistors, *Nat. Nanotechnol.*, **18**, 1044–1050, (2023)
5. KH Kim, I. Karpov, R. H. Olsson, D. Jariwala, Wurtzite and fluorite ferroelectric materials for electronic memory, *Nat. Nanotechnol.*, **18**, 422–441, (2023)
6. S. Fichtner, N. Wolff, F. Lofink, L. Kienle, B. Wagner, AlScN: A III-V semiconductor based ferroelectric., *J. Appl. Phys.*, **125**, 11, (2019)
7. S. Clima, C. Pashartis, J. Bizindavyi, S. R. C. McMitchell, M. Houssa, J. V. Houdt, G. Pourtois, Strain and ferroelectricity in wurtzite $Sc_xAl_{1-x}N$ materials., *Appl. Phys. Lett.*, **119**, 17, (2021)
8. Y. Zhang, Q. Zhu, B. Tian, C. Duan , New-Generation Ferroelectric AlScN Materials, *Nano-Micro Lett.*, **16**, 227, (2024)
9. J. X. Zheng, M. M. A. Fiagbenu, G. Esteves, P. Musavigharavi, A. Gunda, D. Jariwala, E. A. Stach, R. H. Olsson, Ferroelectric behavior of sputter deposited $Al_{0.72}Sc_{0.28}N$ approaching 5 nm thickness., *Appl. Phys. Lett.*, **122**, 22, (2023)
10. D. K. Pradhan, D. C. Moore, G. Kim, Y. He, P. Musavigharavi, KH Kim, N. Sharma, Z. Han, X. Du, V. S. Puli, E. A. Stach, W. J. Kennedy, N. R. Glavin, R. H. Olsson, D. Jariwala, A scalable ferroelectric non-volatile memory operating at 600 °C, *Nat Electron*, **7**, 348–355, (2024)
11. SM Chen, T. Hoshii, H. Wakabayashi, K. Tsutsui, E. Y. Chang, K. Kakushima, Reactive sputtering of ferroelectric AlScN films with $H_2$ gas flow for endurance improvement, *Jpn. J. Appl. Phys.*, **63**, 03SP45, (2024)

12. KH Kim, S. Song, B. Kim, P. Musavigharavi, N. Trainor, K. Katti, C. Chen, S. Kumari, J. Zheng, J. M. Redwing, E. A. Stach, R. H. Olsson III, D. Jariwala, Tuning Polarity in WSe$_2$/AlScN FeFETs via Contact Engineering, *ACS Nano*, **18**, 4180-4188, (2024)

13. W. Zhu, J. Hayden, F. He, JI Yang, P. Tipsawat, M. D. Hossain, JP Maria, S. Trolier-McKinstry, Strongly temperature dependent ferroelectric switching in AlN, Al$_{1-x}$Sc$_x$N, and Al$_{1-x}$B$_x$N thin films., *Appl. Phys. Lett.*, **119**, 062901., (2021)

14. D. Wang, P. Musavigharavi, J. Zheng, G. Esteves, X. Liu, M. M. A. Fiagbenu, E. A. Stach, D. Jariwala, R. H. Olsson III, Sub-Microsecond Polarization Switching in (Al,Sc)N Ferroelectric Capacitors Grown on Complementary Metal–Oxide–Semiconductor-Compatible Aluminum Electrodes., *Phys. Status Solidi RRL*, **15**: 2000575, (2021)

15. A. Grigoriev, M. M. Azad, J. McCampbell, Ultrafast electrical measurements of polarization dynamics in ferroelectric thin-film capacitors, *Rev. Sci. Instrum.*, **82**, 12, (2011)

16. Z. Tang, G. Esteves, R. H. Olsson, Sub-quarter micrometer periodically poled Al$_{0.68}$Sc$_{0.32}$N for ultra-wideband photonics and acoustic devices., *J. Appl. Phys.*, **134**, 11, (2023)

17. R. Guido, H. Lu, P. D. Lomenzo, T. Mikolajick, A. Gruverman, U. Schroeder, Kinetics of N- to M-Polar Switching in Ferroelectric Al$_{1-x}$Sc$_x$N Capacitors., *Adv. Sci.*, **11**, 2308797., (2024)

18. CW Lee, K. Yazawa, A. Zakutayev, G. L. Brennecka, P. Gorai, Switching it up: New mechanisms revealed in wurtzite-type ferroelectrics, *Sci. Adv.*, **10**, eadl0848, (2024)

19. R. Guido, X. Wang, B. Xu, R. Alcala, T. Mikolajick, U. Schroeder, P. D. Lomenzo, Ferroelectric Al$_{0.85}$Sc$_{0.15}$N and Hf$_{0.5}$Zr$_{0.5}$O$_2$ Domain Switching Dynamics, *ACS Applied Materials & Interfaces*, **16**, 42415-42425, (2024)

20. KH Kim, Z. Han, Y. Zhang, P. Musavigharavi, J. Zheng, D. K. Pradhan, E. A. Stach, R. H. Olsson III, D. Jariwala, Multistate, Ultrathin, Back-End-of-Line-Compatible AlScN Ferroelectric Diodes, *ACS Nano*, **18**, 15925-15934, (2024)

21. H. Mulaosmanovic, E. T. Breyer, T. Mikolajick, S. Slesazeck, Reconfigurable frequency multiplication with a ferroelectric transistor, *Nat Electron*, **3**, 391–397, (2020)

22. S. Oh, H. Hwang, I. K. Yoo, Ferroelectric materials for neuromorphic computing., *APL Mater.*, **7**, 091109, (2019)

23. X. Liu, J. Ting, Y. He, M. M. A. Fiagbenu, J. Zheng, D. Wang, J. Frost, P. Musavigharavi, G. Esteves, K. Kisslinger, S. B. Anantharaman, E. A. Stach, R. H. Olsson III, D. Jariwala, Reconfigurable Compute-In-Memory on Field-Programmable Ferroelectric Diodes, *Nano Letters*, **22**, 7690-7698, (2022)

24. V. Gund, B. Davaji, H. Lee, J. Casamento, H. G. Xing, D. Jena, A. Lal, Towards Realizing the Low-Coercive Field Operation of Sputtered Ferroelectric Sc$_x$Al$_{1-x}$N, *2021 21st International*

*Conference on Solid-State Sensors, Actuators and Microsystems (Transducers), Orlando, FL, USA,* 1064-1067, (2021)

25. D. Wang, P. Wang, B. Wang, Z. Mi, Fully epitaxial ferroelectric ScGaN grown on GaN by molecular beam epitaxy., *Appl. Phys. Lett.*, **119**, 111902, (2021)

26. R. Cao, B. Song, D. Shang, Y. Yang, Q. Luo, S. Wu, Y. Li, Y. Wang, H. Lv, Q. Liu, M. Liu, Improvement of Endurance in HZO-Based Ferroelectric Capacitor Using Ru Electrode, *IEEE Electron Device Letters*, **40**, 1744-1747, (2019)

27. M. Lederer, R. Olivo, D. Lehninger, S. Abdulazhanov, T. Kämpfe, S. Kirbach, C. Mart, K. Seidel, L. M. Eng, On the Origin of Wake-Up and Antiferroelectric-Like Behavior in Ferroelectric Hafnium Oxide, *Phys. Status Solidi RRL*, **15**, 2100086, (2021)

28. S. Ray, AN INTRODUCTION TO HIGH VOLTAGE ENGINEERING, 2nd Edition (PHI Learning Ltd., 2013)

29. J. P. Joule, P. M. Roget, On the production of heat by voltaic electricity, *Proc. R. Soc. Lond.*, **4**: 280–282, (1843)

30. A. Christou, Electromigration and Electronic Device Degradation (Wiley, 1994)

31. B. Max, M. Hoffmann, S. Slesazeck, T. Mikolajick, Direct Correlation of Ferroelectric Properties and Memory Characteristics in Ferroelectric Tunnel Junctions, *IEEE Journal of the Electron Devices Society*, **7**, 1175-1181, (2019)

32. P. Wang, D. Wang, N. M. Vu, T. Chiang, J. T. Heron, Z. Mi, Fully epitaxial ferroelectric ScAlN grown by molecular beam epitaxy, *Appl. Phys. Lett.,* **118**, 223504, (2021)

33. SL Tsai, T. Hoshii, H. Wakabayashi, K. Tsutsui, TK Chung, E. Y. Chang, K. Kakushima, Field cycling behavior and breakdown mechanism of ferroelectric $Al_{0.78}Sc_{0.22}N$ films, *Jpn. J. Appl. Phys.*, **61**, SJ1005, (2022)

34. M. Pešić, F. P. G. Fengler, L. Larcher, A. Padovani, T. Schenk, E. D. Grimley, X. Sang, J. M. LeBeau, S. Slesazeck, U. Schroeder, T. Mikolajick, Physical Mechanisms behind the Field-Cycling Behavior of $HfO_2$-Based Ferroelectric Capacitors., *Adv. Funct. Mater.*, **26**, 4601-4612., (2016)

35. M, Pešić, C, Künneth, M, Hoffmann, H. Mulaosmanovic, S. Müller, E. T. Breyer, U. Schroeder, A. Kersch, T. Mikolajick, St. Slesazeck, A computational study of hafnia-based ferroelectric memories: from ab initio via physical modeling to circuit models of ferroelectric device, *J Comput Electron*, **16**, 1236–1256, (2017)

36. F. He, W. Zhu, J. Hayden, J. Casamento, Q. Tran, K. Kang, Y. Song, B.l Akkopru-Akgun, J. I. Yang, P. Tipsawat, G. Brennecka, S. Choi, T. N. Jackson, JP Maria, S. Trolier-McKinstry, Frequency dependence of wake-up and fatigue characteristics in ferroelectric $Al_{0.93}B_{0.07}N$ thin films, *Acta Materialia*, **266**, 119678, (2024)

37. R. Soni, A. Petraru, P. Meuffels, O. Vavra, M. Ziegler, S. K. Kim, D. S. Jeong, N. A. Pertsev, H. Kohlstedt, Giant electrode effect on tunnelling electroresistance in ferroelectric tunnel junctions. ,*Nat Commun*, **5**, 5414, (2014)

38. S. Zhang, H. Hao, R. Huang, M. Cao, Z. Yao, H. Liu, Enhanced breakdown strength and polarization behavior in relaxor ferroelectric films via bidirectional design of defect engineering and heterogeneous interface construction, *J. Mater. Chem. C*, (2025)

39. L. Chen, C. Liu, H. K. Lee, B. Varghese, R. W. Ip, M. Li, Z. J. Quek, Y. Hong, W. Wang, W. Song, H. Lin, Demonstration of 10 nm Ferroelectric $Al_{0.7}Sc_{0.3}N$-Based Capacitors for Enabling Selector-Free Memory Array, *Materials*, **17**, 627., (2024)

40. K. D. Kim, Y. B. Lee, S. H. Lee, I. S. Lee, S. K. Ryoo, S. Byun, J. H. Lee, H. Kim, H. W. Park, C. S. Hwang, Evolution of the Ferroelectric Properties of AlScN Film by Electrical Cycling with an Inhomogeneous Field Distribution, *Adv. Electron. Mater.*, **0**, 2201142., (2023)

41. D. Drury, K. Yazawa, A. Zakutayev, B. Hanrahan, G. Brennecka, High-Temperature Ferroelectric Behavior of $Al_{0.7}Sc_{0.3}N$, *Micromachines*, **13**(6), 887, (2022)

42. K. D. Kim, Y. B. Lee, S. H. Lee, I. S. Lee, S. K. Ryoo, S. Y. Byun, J. H. Lee, C. S. Hwang, Impact of operation voltage and $NH_3$ annealing on the fatigue characteristics of ferroelectric AlScN thin films grown by sputtering, *Nanoscale*, **15**, 16390-16402, (2023)

43. S. K. Ryoo, K. D. Kim, H. W. Park, Y. B. Lee, S. H. Lee, I. S. Lee, S. Byun, D. Shim, J. H. Lee, H. Kim, Y. H. Jang, M. H. Park, C. S. Hwang, Investigation of Optimum Deposition Conditions of Radio Frequency Reactive Magnetron Sputtering of $Al_{0.7}Sc_{0.3}N$ Film with Thickness down to 20 nm., *Adv. Electron. Mater.*, **8**, 2200726., (2022)

44. Yunfei He, Shangyi Chen, Merrilyn Mercy Adzo Fiagbenu, Chloe Leblanc, Pariasadat Musavigharavi, Gwangwoo Kim, Xingyu Du, Jiazheng Chen, Xiwen Liu, Eric A. Stach, Roy H. Olsson, Deep Jariwala, Metal-ferroelectric AlScN-semiconductor memory devices on SiC wafers, Appl. Phys. Lett., 123 (12): 122901., (2023)

45. Roberto Guido, Thomas Mikolajick, Uwe Schroeder, Patrick D. Lomenzo, Role of Defects in the Breakdown Phenomenon of $Al_{1-x}Sc_xN$: From Ferroelectric to Filamentary Resistive Switching, Nano Lett., 23, 7213-7220, (2023)

46. H. J. Joo, S. S. Yoon, S. Y. Oh, Y. Lim, G. H. Lee, G. Yoo, Temperature-Dependent Ferroelectric Behaviors of AlScN-Based Ferroelectric Capacitors with a Thin $HfO_2$ Interlayer for Improved Endurance and Leakage Current, *Electronics*, **13**(22), 4515, (2024)

47. J. Casamento, F. He, C. Skidmore, J. Hayden, J. Nordlander, J. M. Redwing, S. Trolier-McKinstry, JP Maria, Ferroelectric $Al_{1-x}B_xN$–GaN heterostructures, *Appl. Phys. Lett.*, **124** (14): 142101, (2024)

48. M. Uehara, K. Hirata, Y. Nakamura, S. A. Anggraini, K. Okamoto, H. Yamada, H. Funakubo, M. Akiyama, Excellent piezoelectric and ferroelectric properties of $Sc_xGa_{1-x}N$ alloy with high Sc concentration, *APL Mater*, **12** (12): 121102, (2024)

49. X. Liu, D. Zhou, Y. Guan, S. Li, F. Cao, X. Dong, Endurance properties of silicon-doped hafnium oxide ferroelectric and antiferroelectric-like thin films: A comparative study and prediction, *Acta Materialia*, **154**, 190-198, (2018)

50. Z. Fu, S. Cao, H. Zheng, J. Luo, Q. Huang, R. Huang, Hafnia-Based High-Disturbance-Immune and Selector-Free Cross-Point FeRAM, *IEEE Transactions on Electron Devices*, vol. **71**, no. 5, pp. 3358-3364, (2024)

51. J. Li, H. Wang, X. Du, Z. Luo, Y.n Wang, W. Bai, X. Su, S. Shen, Y.i Yin, X. Li, High endurance ($>10^{12}$) via optimized polarization switching ratio for $Hf_{0.5}Zr_{0.5}O_2$-based FeRAM, *Appl. Phys. Lett.*, **122** (8): 082901, (2023)

52. M. I. Popovici, J. Bizindavyi, P. Favia, S. Clima, Md. N. K. Alam, R. K. Ramachandran, A. M. Walke, U. Celano, A. Leonhardt, S. Mukherjee, O. Richard, A. Illiberi, M. Givens, R. Delhougne, J. V. Houdt, G. S. Kar, High performance La-doped HZO based ferroelectric capacitors by interfacial engineering, *International Electron Devices Meeting (IEDM)*, 6.4.1-6.4.4, (2022)

53. YD Lin, PC Yeh, JY Dai, JW Su, HH Huang, CY Cho, YT Tang, TH Hou, SS Sheu, WC Lo, SC Chang , Highly Reliable, Scalable, and High-Yield $HfZrO_x$ FRAM by Barrier Layer Engineering and Post-Metal Annealing, *International Electron Devices Meeting (IEDM)*, 32.1.1-32.1.4, (2022)

54. A. G. Chernikova, M. G. Kozodaev, D. V. Negrov, E. V. Korostylev, M. H. Park, U. Schroeder, C. S. Hwang, A. M. Markeev, Improved Ferroelectric Switching Endurance of La-Doped $Hf_{0.5}Zr_{0.5}O_2$ Thin Films, *ACS Applied Materials & Interfaces*, **10** (3), 2701-2708, (2018)

55. N. Ramaswamy, A. Calderoni, J. Zahurak, G. Servalli, A. Chavan, S. Chhajed, M. Balakrishnan, M. Fischer, M. Hollander, D. P. Ettisserry, A. Liao, K. Karda, M. Jerry, M. Mariani, A. Visconti, B. R. Cook, B. D. Cook, D. Mills, A. Torsi, C. Mouli, E. Byers, M. Helm, S. Pawlowski, S. Shiratake, N. Chandrasekaran, NVDRAM: A 32Gb Dual Layer 3D Stacked Non-volatile Ferroelectric Memory with Near-DRAM Performance for Demanding AI Workloads, *International Electron Devices Meeting (IEDM)*, 1-4, (2023)

56. H. Joh; T. Jung; S. Jeon, Stress Engineering as a Strategy to Achieve High Ferroelectricity in Thick Hafnia Using Interlayer, *IEEE Transactions on Electron Devices*, vol. **68**, no. 5, pp. 2538-2542, (2021)

57. R. Moazzami; C. Hu; W. H. Shepherd, Endurance properties of ferroelectric PZT thin films, *International Technical Digest on Electron Devices*, pp. 417-420, (1990)

58. D. C. Yoo, B. J. Bae, J. E. Lim, D. H. Im, S. O. Park, S. H. Kim, U. I. Chung, J. T. Moon, B. I. Ryu, Highly Reliable 50nm-thick PZT Capacitor and Low Voltage FRAM Device using Ir/SrRuO$_3$/MOCVD PZT Capacitor Technology, *Digest of Technical Papers. 2005 Symposium on VLSI Technology*, pp. 100-101, (2005)

59. K. Kim, S. Lee, Integration of lead zirconium titanate thin films for high density ferroelectric random access memory, *J. Appl. Phys.*, **100** (5): 051604, (2006)

60. N. K. Karan, R. Thomas, S. P. Pavunny, J. J. Saavedra-Arias, N. M. Murari, R. S. Katiyar, Preferential grain growth and improved fatigue endurance in Sr substituted PZT thin films on Pt(111)/TiO$_x$/SiO$_2$/Si substrates, *Journal of Alloys and Compounds*, Volume **482**, Issues 1–2, Pages 253-255, (2009)

61. H. Zhu, S. Yamamoto, J. Matsui, T. Miyashita, M. Mitsuishi, Ferroelectricity of poly(vinylidene fluoride) homopolymer Langmuir–Blodgett nanofilms, *J. Mater. Chem. C*, **2**, 6727, (2014)

62. JW Yoon, SMYoon, H. Ishiwara, Improvement in Ferroelectric Fatigue Endurance of Poly(methyl metacrylate)-Blended Poly(vinylidene fluoride–trifluoroethylene), *Jpn. J. Appl. Phys.*, **49** 030201, (2010)

63. D. Zhao, I. Katsouras, M. Li, K. Asadi, J. Tsurumi, G. Glasser, J. Takeya, P. W. M. Blom, D. M. de Leeuw, Polarization fatigue of organic ferroelectric capacitors, *Sci Rep*, **4**, 5075, (2014)

64. Y. Jiang, E. Parsonnet, A. Qualls, W. Zhao, S. Susarla, D. Pesquera, A. Dasgupta, M. Acharya, H. Zhang, T. Gosavi, C.-C. Lin, D. E. Nikonov, H. Li, I. A. Young, R. Ramesh, L. W. Martin, Enabling ultra-low-voltage switching in BaTiO$_3$, *Nat. Mater.*, **21**, 779–785, (2022)

65. J. Zhai, H. Chen, Ferroelectric properties of Bi$_{3.25}$La$_{0.75}$Ti$_3$O$_{12}$ thin films grown on the highly oriented LaNiO$_3$ buffered Pt/Ti/SiO$_2$/Si substrates, *Appl. Phys. Lett.*, **82** (3): 442–444, (2003)

66. A. Haque, H. Jason D'Souza, S. K. Parate, R. S. Sandilya, S. Raghavan, P. Nukala, Heterogeneous Integration of High Endurance Ferroelectric and Piezoelectric Epitaxial BaTiO$_3$ Devices on Si, *Adv. Funct. Mater.*, **35**, 2413515, (2024)

67. M. Scigaj, N. Dix, J. Gázquez, M. Varela, I. Fina, N. Domingo, G. Herranz, V. Skumryev, J. Fontcuberta, F. Sánchez, Monolithic integration of room-temperature multifunctional BaTiO$_3$-CoFe$_2$O$_4$ epitaxial heterostructures on Si(001), *Sci Rep*, **6**, 31870, (2016)

68. M. Park, J. Y. Yang, M. J. Yeom, B. Bae, Y. Baek, G. Yoo, K. Lee, An artificial neuromuscular junction for enhanced reflexes and oculomotor dynamics based on a ferroelectric CuInP$_2$S$_6$/GaN HEMT, *Sci. Adv.*, **9**, eadh9889, (2023)

69. Z. Zhou, S. Wang, Z. Zhou, Y. Hu, Q. Li, J. Xue, Z. Feng, Q. Yan, Z. Luo, Y. Weng, R. Tang, X. Su, F. Zheng, K. Okamoto, H. Funakubo, L. Kang, L. Fang, L. You, Unconventional polarization fatigue in van der Waals layered ferroelectric ionic conductor $CuInP_2S_6$., *Nat Commun*, **14**, 8254, (2023)

70. J. Celinska, V. Joshi, S. Narayan, L. McMillan, C. Paz de Araujo, Effects of scaling the film thickness on the ferroelectric properties of $SrBi_2Ta_2O_9$ ultra thin films., *Appl. Phys. Lett.*, **82** (22): 3937–3939, (2003)

71. S. Sakai, R. Ilangovan, Metal-ferroelectric-insulator-semiconductor memory FET with long retention and high endurance, *IEEE Electron Device Letters*, vol. **25**, no. 6, pp. 369-371, (2004)

Supplementary Information for

# Write Cycling Endurance Exceeding $10^{10}$ in Sub-50 nm Ferroelectric AlScN

Hyunmin Cho[1]†, Yubo Wang[1], Chloe Leblanc[1], Yinuo Zhang[1], Yunfei He[1], Zirun Han[1], Roy H. Olsson III[1]*, Deep Jariwala[1]*

[1]Department of Electrical and Systems Engineering, University of Pennsylvania, Philadelphia, Pennsylvania 19104, USA

*Corresponding author: Deep Jariwala (email: dmj@seas.upenn.edu), Roy H. Olsson III (email: rolsson@seas.upenn.edu)

**Supplementary Figure S1-S15**

**Supplementary Information S1-S3**

**Supplementary References**

# Supplementary Figures

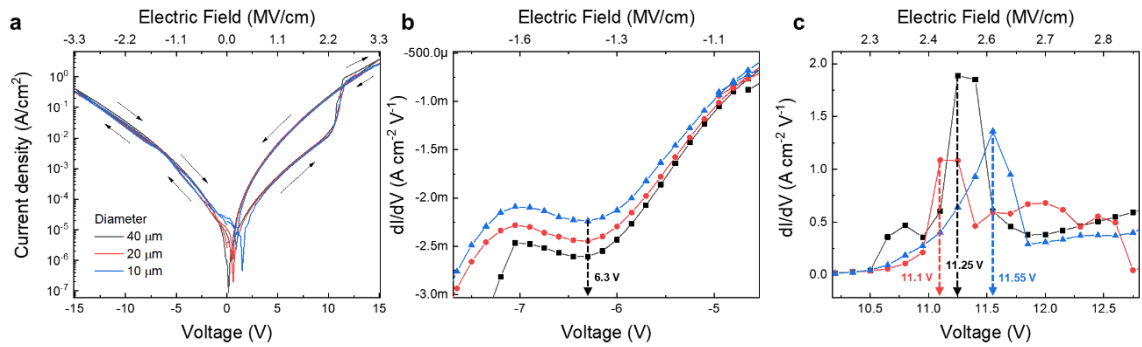

**Supplementary Figure S1 | Quasi-direct current density versus voltage (DC-IV) characteristics.** **(a)** The y-axis represents the current density, calculated as the current divided by the sample area for different diameters. **(b)** and **(c)** show the first derivative of the current density with respect to voltage, which is used to extract the coercive voltage ($V_C$).

As shown in **Figure S1a**, no significant variation was observed among different diameters, indicating high reliability with minimal dependence on capacitor area, consistent with the results in **Figure 1d** and **e**. To determine the $V_C$ from the plots in **Figure S1a**, we calculated the first derivative of the current density with respect to voltage. The $V_C$ was identified as the voltage at which this derivative reaches a local maximum, corresponding to the point where the rate of change in current density is maximized. **Figure S1b** and **S1c** represent the current responses under negative and positive voltage applications, respectively, from which the $V_C$ were extracted for each case. Moreover, the negative coercive voltage remains unchanged regardless of diameter. Although the positive coercive voltage exhibits some variation, these differences arise from measurement errors induced by high leakage.

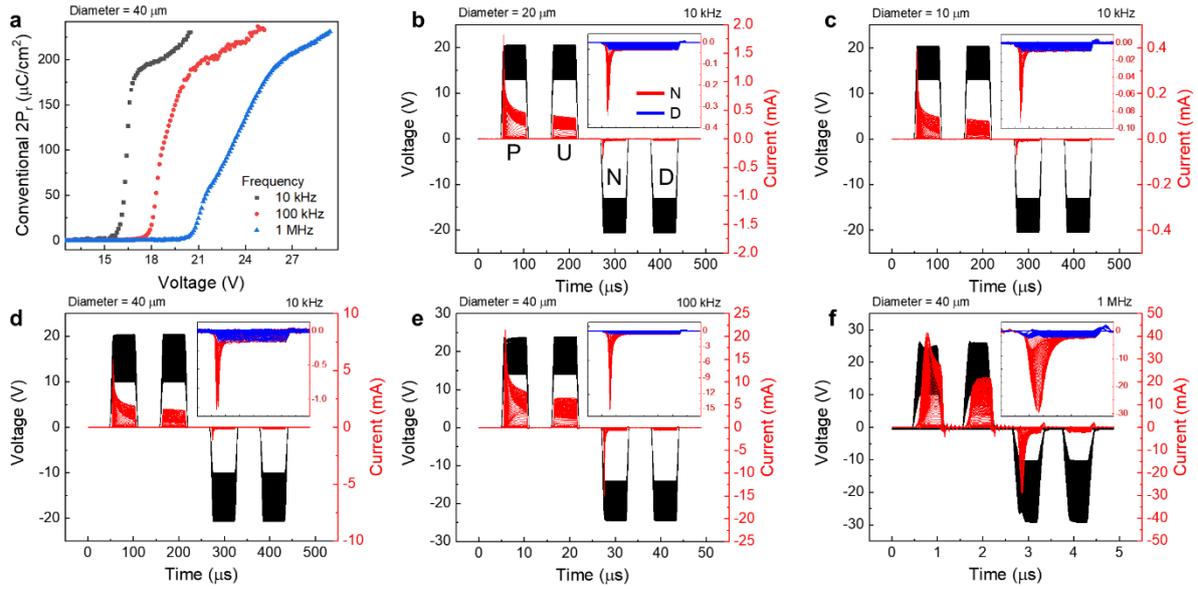

**Supplementary Figure S2 | PUND measurements under various conditions. (a)** Evolution of conventional 2P$_r$ values as a function of applied voltage across varying PUND frequencies, with voltage increments of 0.1 V step. **(b–f)** show detailed current and voltage characteristics across different capacitor diameters and frequency settings. The insets in each plot highlight the difference between N and D pulses by overlaying their respective current responses. The red and blue curves represent the current responses measured when applying the N and D pulses, respectively. **(b–d)** present detailed PUND results explaining each graph in **Figure 1e**, providing a clearer interpretation of the observed responses. Similarly, **(d–f)** present detailed PUND results corresponding to **Figure S1a**, providing a clearer view of frequency-dependent behavior in 40 μm diameter capacitors.

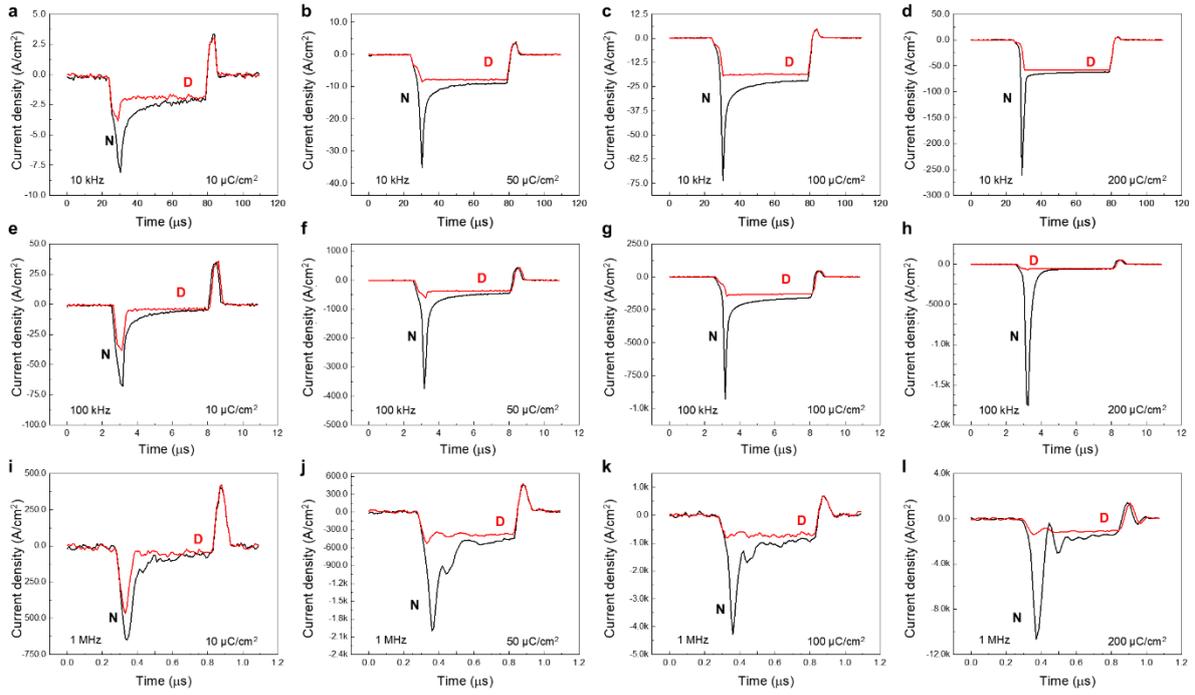

**Supplementary Figure S3 | Overlaid current density responses correspond to the difference between N and D voltage pulses during the endurance test.** These results are based on PUND measurements taken before leakage started during the endurance test. The data is arranged with increasing frequency in the lower rows and increasing conventional $2P_r$ in the rightward columns. A clear trend shows that higher frequencies lead to an increase in the peak current density, requiring a higher operating voltage to maintain the consistent conventional $2P_r$. Additionally, within the same frequency (same row), moving toward higher conventional $2P_r$ (rightward columns) results in larger current density peaks. This is because, achieving a higher conventional $2P_r$ at a given frequency requires a larger applied voltage, leading to a higher current response. In all cases, the current responses consistently validate the ferroelectric characteristics, confirming the presence of partial polarization in AlScN.

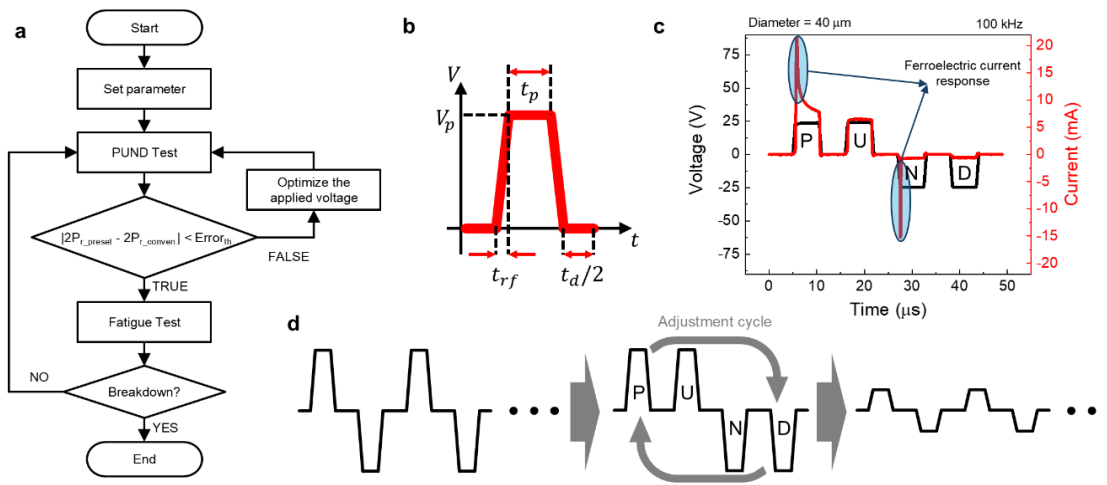

**Supplementary Figure S4 | Algorithm for self-adjusting voltage. (a)** Overall flowchart for the test. **(b)** Single-pulse configuration with relevant time and voltage parameters, including voltage pulse amplitude ($V_p$), pulse width ($t_p$), rise/fall time ($t_{rf}$), and delay time ($t_d$). The reason a single pulse only has $t_d/2$ instead of $t_d$ is that the following pulse contributes $t_d/2$, ensuring the full delay time. The frequency is defined as $1/(2 \times t_p)$. **(c)** Detailed PUND measurement results. The blue ellipses in the plots indicate the ferroelectric current response. Other current responses are attributed to leakage and RC delay currents. To isolate the ferroelectric response in PUND measurements, the charge calculated from P and N is subtracted by the charge from U and D (P−U and N−D). The extracted values are then normalized by the capacitor area to obtain the charge density. **(d)** illustrates the pulse train configuration for the fatigue and PUND tests. The first pulse train alternates between positive and negative identical pulses applied to the capacitor. After a certain number of cycles, a PUND measurement is conducted to maintain a stable conventional $2P_r$ by adjusting the applied voltage amplitude. Once the PUND adjustment is completed, the next fatigue pulse train proceeds with the adjusted voltage amplitude for a specified number of cycles.

The algorithm which is performed during the adjustment cycle dynamically adjusts the learning rate for stable voltage regulation. Using real-time feedback, it fine-tunes voltage to align with a target value ($2P_{r\_preset}$), minimizing oscillations and ensuring rapid stabilization. The process begins by analyzing the 2$^{nd}$ derivative of voltage with respect to error ($|2P_{r\_preset} - 2P_{r\_conven.}|$). A negative curvature is preferred to maintain stability in AlScN phase transitions. However, early-phase fluctuations may cause abrupt error reversals, requiring aggressive learning rate adjustments to prevent unnecessary oscillations. Once refined, the algorithm modifies the voltage using a scaled hyperbolic tangent (tanh) function, ensuring smooth, controlled adjustments. The tanh function prevents extreme corrections while maintaining sensitivity to error trends. This cycle repeats until stability is achieved. A final assessment ensures consistency, eliminating fluctuations.

PUND measurements were conducted to quantify the amount of polarization switching before and after fatigue testing. However, the reliability of this method becomes questionable when operating in the partial switching regime. During the P and U steps, polarization switching occurs twice before the application of the N and D pulses, which differs from the actual fatigue condition involving a single switching event per cycle. The switched polarization during the N and D pulses strongly depends on the initial state defined by the preceding P and U pulses. Since this state is different from the one in fatigue testing, particularly under partial switching, the extracted values from PUND measurements are not directly representative of the true fatigue-induced switching behavior. Thus, PUND cannot accurately resolve the switching characteristics in this regime. Despite these limitations, PUND was employed in this study to maintain consistency with prior reports and enable a comparative analysis within this work. A more accurate evaluation of partial polarization switching will require further methodological refinement and validation.

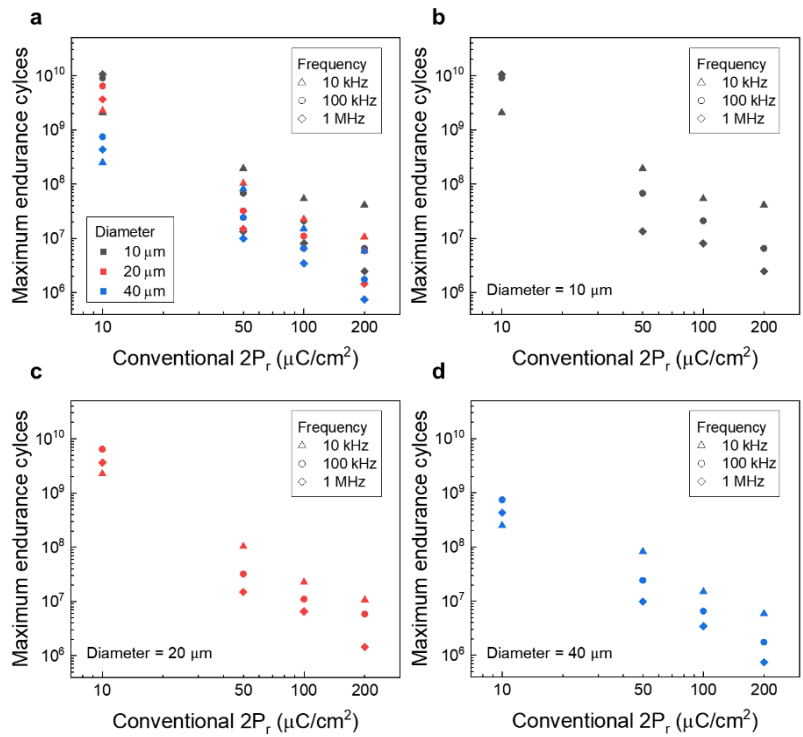

**Supplementary Figure S5 | Reorganized plots representing relationships between conventional 2P$_r$ and endurance (a)** Comprehensive dataset covering all conditions. Subdivided plots for different diameters: **(b)** 10 μm, **(c)** 20 μm, and **(d)** 40 μm.

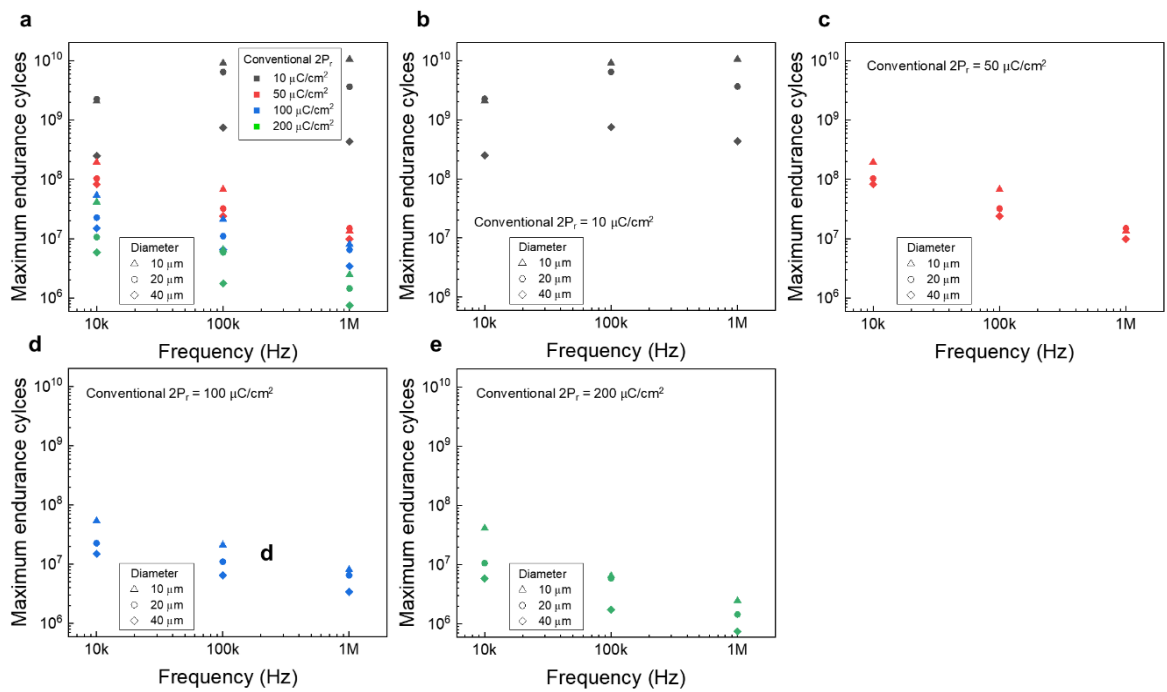

**Supplementary Figure S6 | Reorganized plots representing relationships between frequency and endurance (a)** Comprehensive dataset covering all conditions. Subdivided plots for maintained conventional $2P_r$: **(b)** 10 μC/cm², **(c)** 50 μC/cm², **(d)** 100 μC/cm², and **(e)** 200 μC/cm².

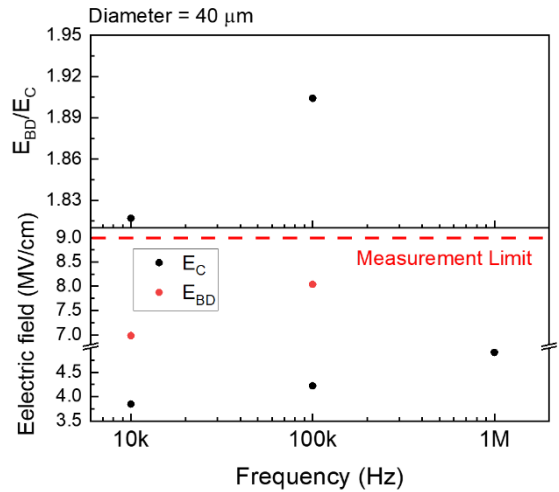

**Supplementary Figure S7 | Frequency Dependence of Coercive and Breakdown Electric Fields.** This plot illustrates the correlation between the operation frequency in PUND measurements and key ferroelectric properties, specifically the coercive electric field and the breakdown electric field (bottom panel), as well as their ratio (top panel). The red dashed line indicates the operational limit of our measurement system. Consequently, obtaining the breakdown voltage for the 1 MHz case is impossible within the constraints of this setup.

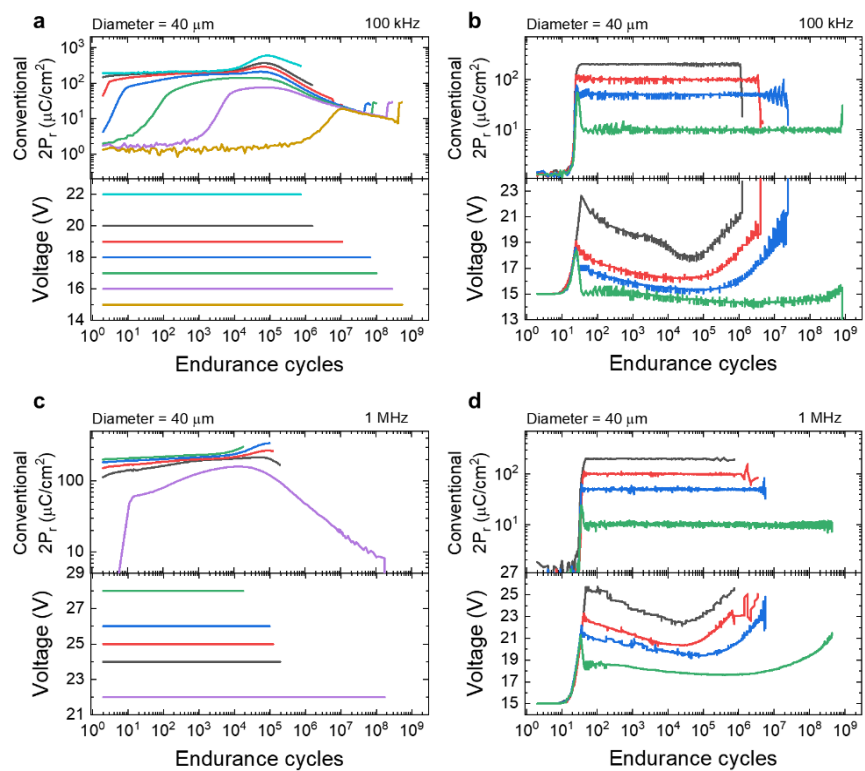

**Supplementary Figure S8 | Endurance test results under various frequency conditions.** Endurance cycles were conducted with **(a)** and **(c)** a constant applied voltage pulse, while **(b)** and **(d)** used an adjusted applied voltage pulse to maintain the conventional $2P_r$ close to preset $2P_r$.

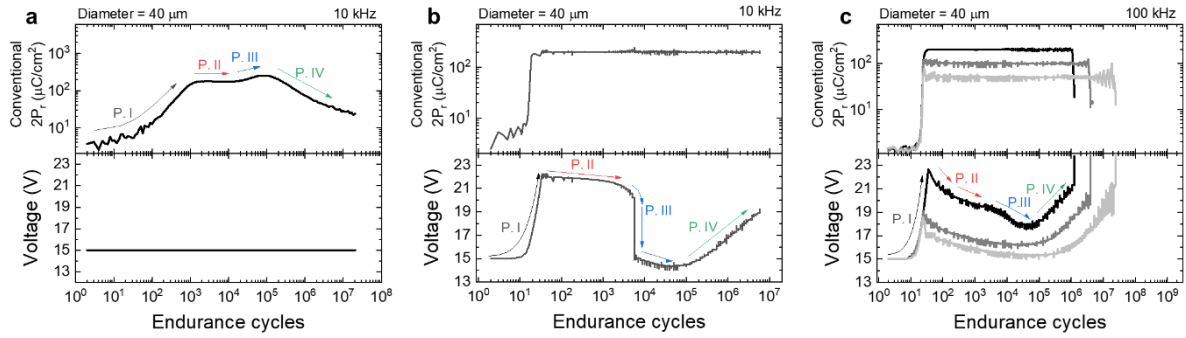

**Supplementary Figure S9 | Phase evolution during endurance test.** Data was obtained from a 40 μm capacitor. **(a)** shows phase evolution for 15 V at 10 kHz. The arrows mark phase changes through variations in $2P_r$. **(b)** adjusts the voltage to keep the conventional $2P_r$ at 200 μC/cm², tracking phase shifts through applied voltage variations. **(c)** shows the results with varying frequency and under conditions that maintain an even lower conventional $2P_r$, ensuring the existence from wake up to reduce degradation effects.

Each phase (P.) in **Figures S9 a–c** indicates distinct polarization phases. A comprehensive analysis is provided in **Supplementary Information S3**. In **Figure S9a**, P. I corresponds to the wake-up process, P. II represents the stabilized ferroelectric state, P. III is dominated by leakage currents, and P. IV signifies the onset of fatigue-induced degradation. In **Figure S9b**, P. I involves voltage modulation to establish the optimal amplitude for the conventional $2P_r$, approaching the preset $2P_r$ value of 200 μC/cm², along with the initial quick wake-up phase. P. II represents the stabilized ferroelectric state, P. III is characterized by increased leakage current contributions, and P. IV marks the progression of degradation. In **Figure S9c**, P. I show the same behavior as in **Figure S9b**. However, in P. II, partial wake-up occurs simultaneously with the stable phase, indicating an extended polarization stabilization period. P. III is characterized by increased leakage current contributions, and P. IV marks the progression of degradation effects. This distinction highlights that for lower preset $2P_r$ conditions, the second phase sustains partial wake-up, leading to the coexistence of the stable and wake-up phases, which is absent in the higher preset $2P_r$ case.

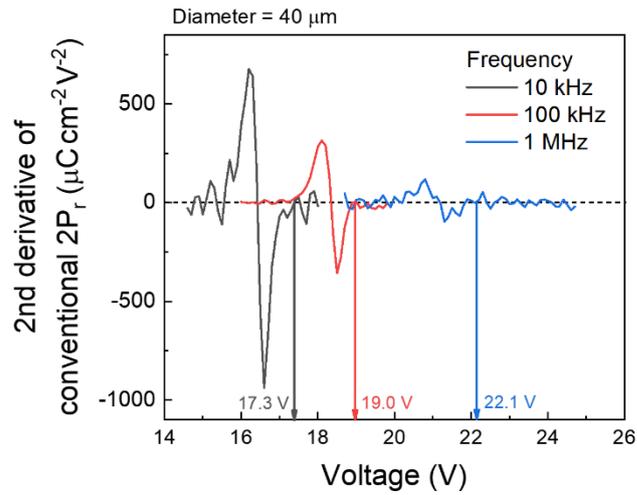

**Supplementary Figure S10 | Extraction of the coercive voltage ($V_c$).** $V_c$ is determined through PUND measurements by analyzing the second derivative of the conventional $2P_r$. The point where this derivative reaches zero signifies polarization saturation. This method enables more meaningful comparisons, making it well-suited for the PUND-based endurance test used in this work.

The second derivative of conventional $2P_r$ decreases as frequency increases, aligning with the trend observed in the slope variation of conventional $2P_r$ against applied voltage, as shown in **Figure S2a**.

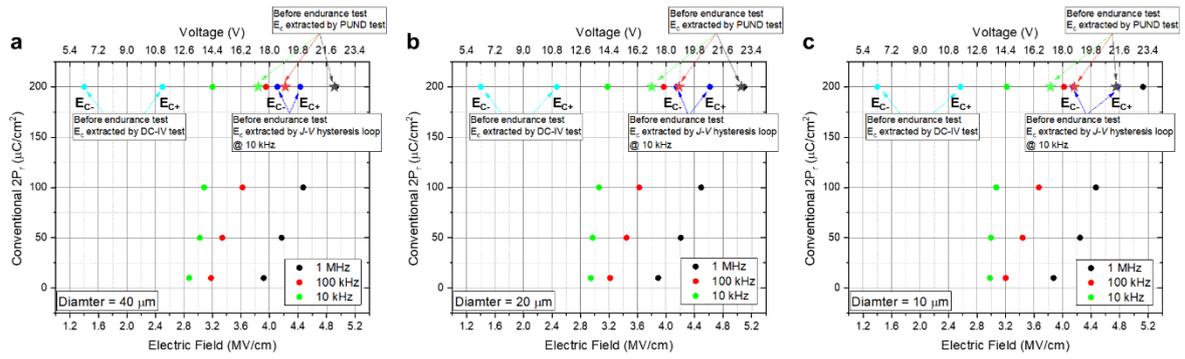

**Supplementary Figure S11 | Conventional 2P$_r$ as a function of applied voltage and electric field for different device sizes. (a)** 40 μm, **(b)** 20 μm, and **(c)** 10 μm diameter devices. Scatter symbols represent different measurement conditions: black, red, and green stars represent conventional 2P$_r$ values obtained before the endurance test using the PUND measurement (as shown in **Figure S10**) at 1 MHz, 100 kHz, and 10 kHz, respectively. Black, red, and green circles represent the adjusted applied voltage before the leakage phase begins at 1 MHz, 100 kHz, and 10 kHz, respectively. Blue circles represent E$_C$ extracted from *J-V* hysteresis loops (**Figure 1e**) at 10 kHz before the endurance test, and cyan circles represent E$_C$ extracted from DC-IV curves (**Figure S1a**). The E$_{C-}$ and E$_{C+}$ indicate results for negative and positive voltage, respectively.

Notably, the change of conventional 2P$_r$ against electric field is slower as frequency increases. This suggests that high frequency enables more precise control over polarization switching. In contrast, at low frequency, the steep slope indicates that partial switching is difficult to control.

Although both E$_{C-}$ and E$_{C+}$ were obtained, E$_{C-}$ should be used for analysis to maintain consistency with our endurance test comparisons. As observed here, the E$_C$ extracted from DC-IV (cyan circle) is much lower than that of the *J-V* hysteresis loop (blue circle) or PUND (green circle). This occurs because the E$_C$ generally decreases with lower frequency, and the applied voltage in the DC-IV measurement, being close to 0 Hz, results in the lowest E$_C$[1-5]. Additionally, the E$_{C-}$ from the *J-V* hysteresis loop (blue circle) and PUND (green star) remains higher than the applied electric field obtained from the endurance test (green circle). This suggests that the wake-up effect lowers the E$_C$.

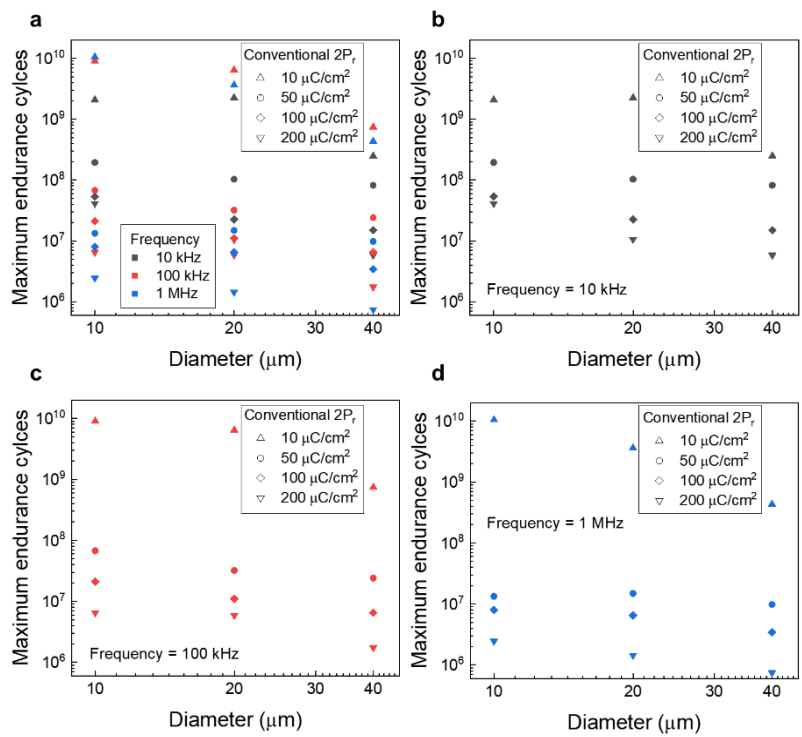

**Supplementary Figure S12 | Reorganized plots representing relationships between diameter and endurance (a)** Comprehensive dataset covering all conditions. Subdivided plots for different frequency: **(b)** 10 kHz, **(c)** 100 kHz, and **(d)** 1 MHz.

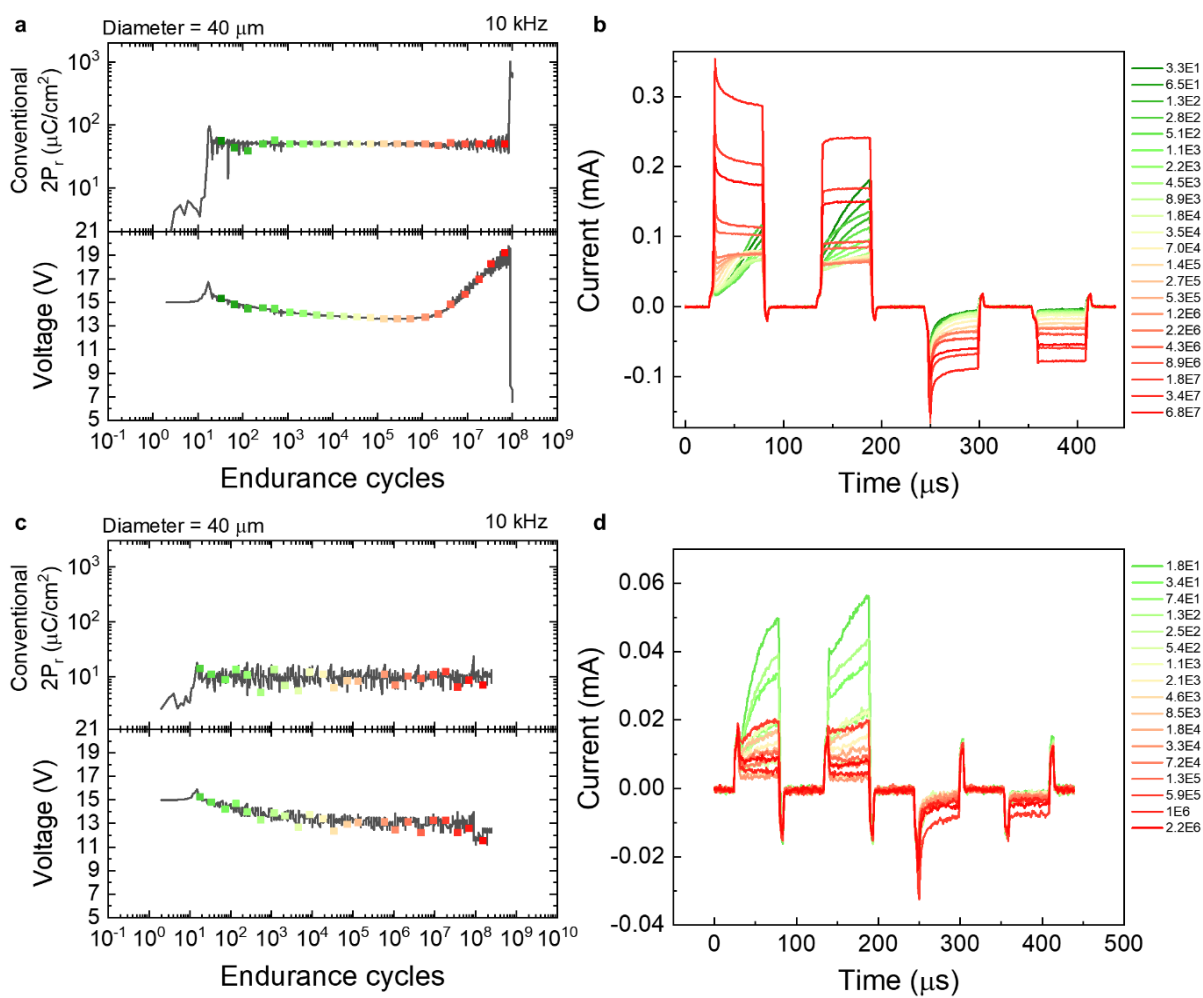

**Supplementary Figure S13 | PUND measurement results obtained from endurance cycling. (a)** and **(c)** Endurance test result of a capacitor with a 40 μm diameter, maintaining a conventional $2P_r$ value of 50 and 10 μC/cm² under a 10 kHz pulse. The rainbow markers correspond to the different plots shown in **(b)** and **(d).** The legend on **(b)** and **(d)** indicates the moment when the PUND measurement was performed.

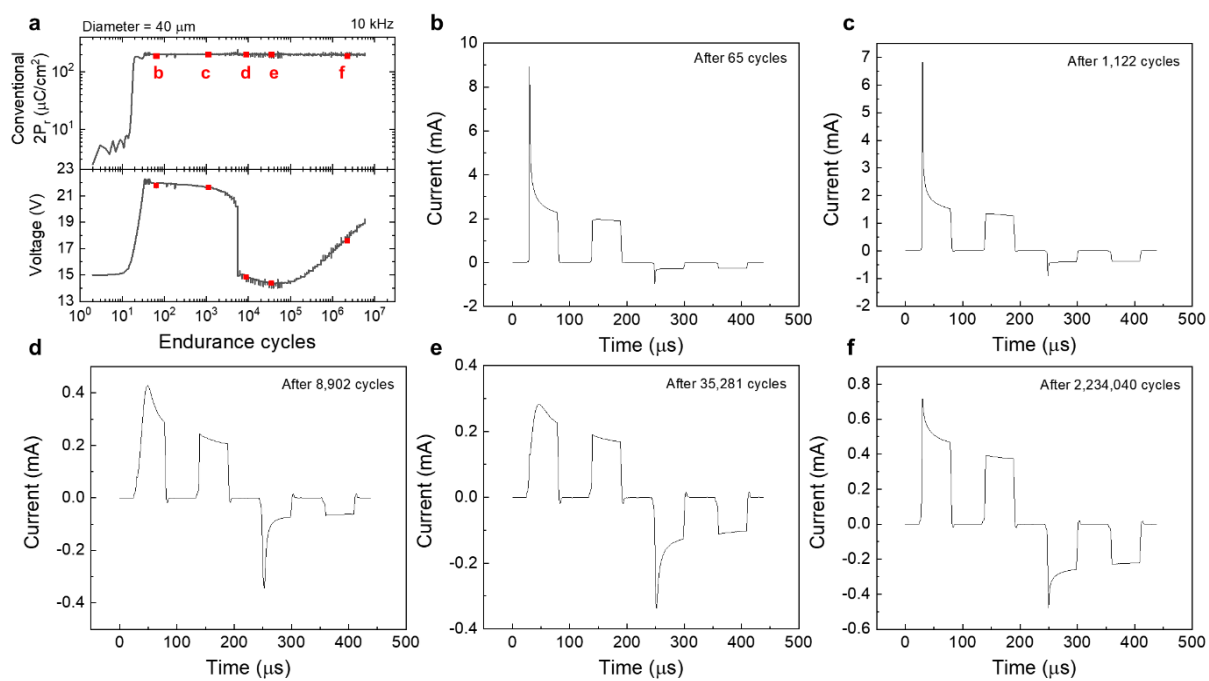

**Supplementary Figure S14 | PUND measurement results obtained from endurance cycling. (a)** Endurance test result of a 40 μm diameter capacitor, maintaining conventional $2P_r$ at 200 μC/cm² under a 10 kHz pulse. The red markers indicate specific cycles corresponding to **(b-f)**, where PUND measurements were conducted to evaluate polarization switching behavior.

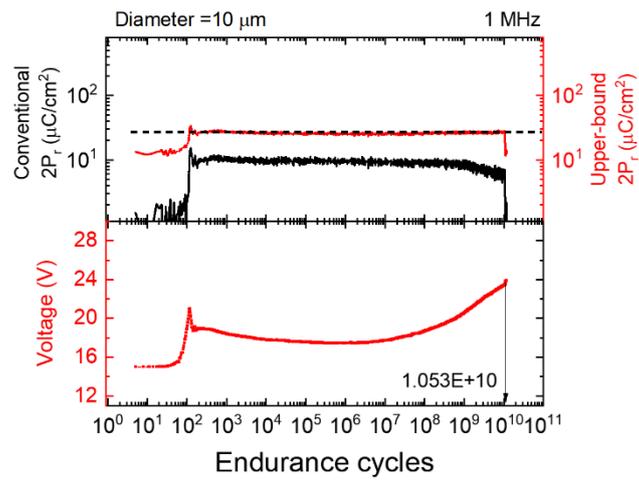

**Supplementary Figure S15 | The best endurance test result.**

**Supplementary Information S1**

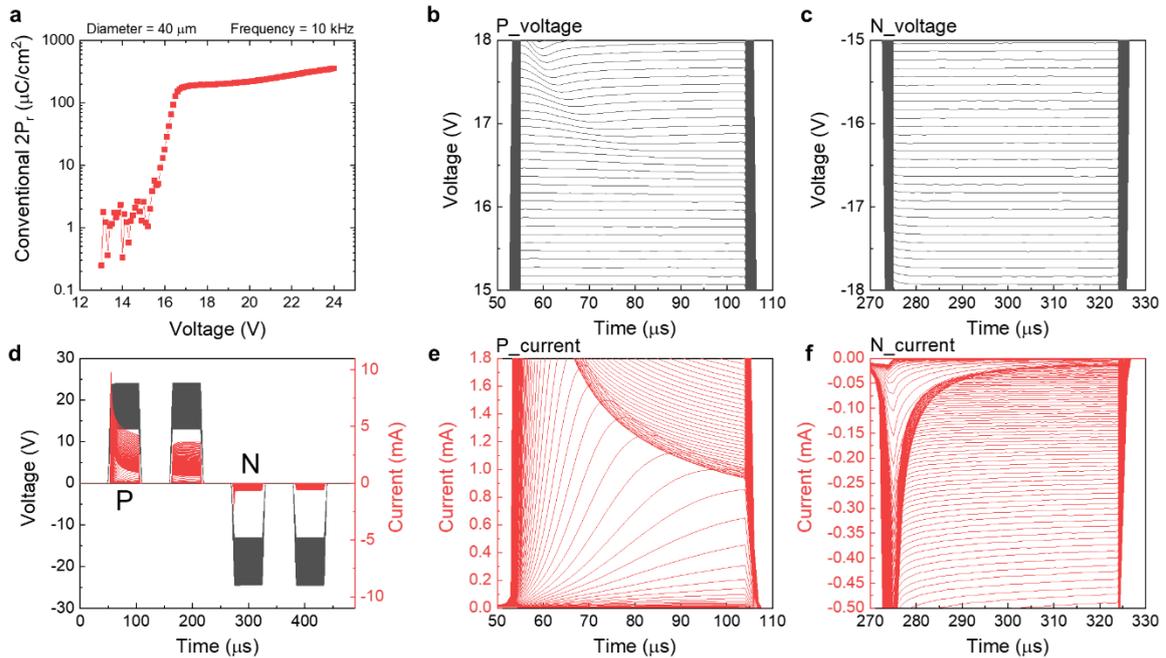

**Supplementary Information Figure S1-1 | PUND measurement results at 10 kHz for a capacitor with a 40 μm diameter.** (a) illustrates conventional $2P_r$ behavior as a function of applied voltage. Voltage response during the (b) P pulse and (c) N pulse. (d) PUND measurement results, including both current and voltage responses over the full-time span. Current response during the (e) P pulse and (f) N pulse.

To deepen the understanding of our AlScN capacitor's partial polarization, we conducted more investigations through the PUND measurements under various conditions. The PUND measurements were conducted systematically by varying the applied voltage pulse in 0.1 V increments and the pulse width. As shown in **Supplementary Information Figure S1-1** and **S1-2,** the observed shift of the current peak to earlier time points, along with its increasing sharpness and magnitude, indicates that such behavior occurs during the polarization switching process. This behavior indicates that domain nucleation and growth occur progressively, with only a fraction of ferroelectric domains switching per pulse instead of undergoing instantaneous collective reversal.

These results are similar to prior studies[6-10]. Their studies demonstrated that the kinetics of ferroelectric switching are strongly influenced by the external field amplitude, pulse duration, and pre-existing domain structures. Guido et al.[9] showed that in AlScN capacitors, polarization switching occurs through domain nucleation and wall motion, with higher electric fields accelerating switching. Yazawa et al.[10] also reported that in wurtzite-structured ferroelectrics, increasing the electric field leads to sharper, larger, and faster current peaks during polarization switching.

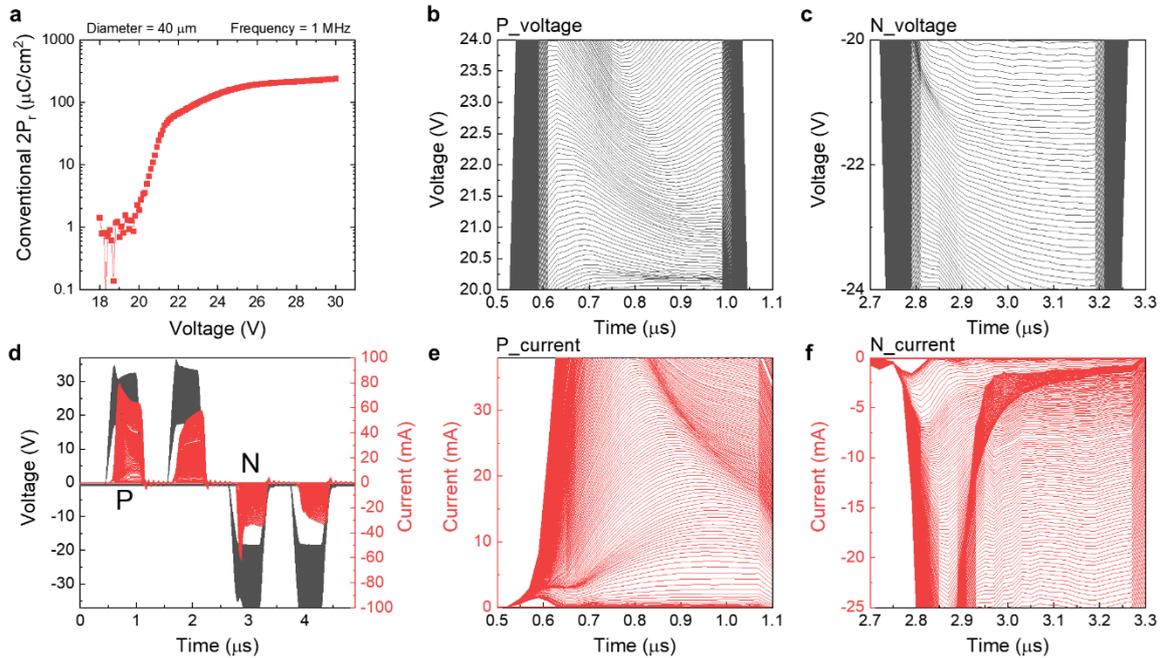

**Supplementary Information Figure S1-2 | PUND measurement results at 1 MHz for a capacitor with a 40 μm diameter.** (a) illustrates conventional $2P_r$ behavior as a function of applied voltage. Voltage response during the (b) P pulse and (c) N pulse. (d) PUND measurement results, including both current and voltage responses over the full-time span. Current response during the (e) P pulse and (f) N pulse.

The increase in the current peak height observed in our measurements provides further evidence of an enhanced switching efficiency as the applied voltage increases. This trend suggests that a greater proportion of ferroelectric domains participate in the switching process at higher fields, leading to an amplified current response. The increase in peak current is likely a consequence of the progressive reduction in activation energy barriers for domain nucleation, as well as the cumulative contribution of switched dipoles. Such behavior is similar to the previous report[7].

**Supplementary Information S2**

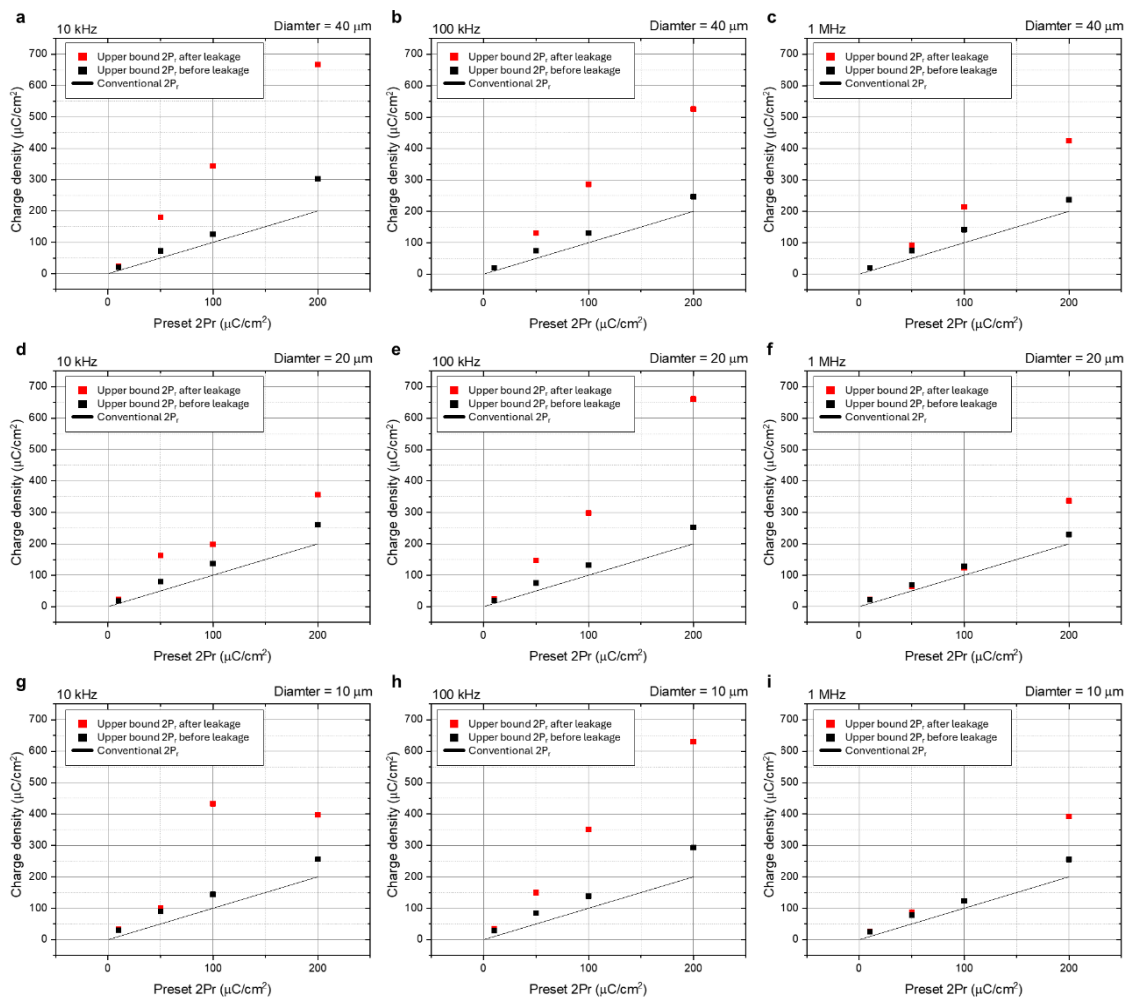

**Supplementary Information Figure S2-1 | Relationships between preset 2P$_r$ and various 2P$_r$ parameters under different conditions.** Each graph contains an endurance test frequency and the area of AlScN capacitors. The line on the graph represents the conventional 2P$_r$, which is the same value as the preset 2P$_r$ due to the operation of a well-maintaining algorithm.

The conventional PUND method is not appropriate to determine partial polarization. Firstly, there is a possibility of underestimation. Conventional 2P$_r$ is calculated using the P-U (subtracting the integrated charge U from P) and the N-D method, both divided by the capacitor area (details in **Supplementary Figure S2**). The first pulse response consists of both leakage (Q$_{leak}$) and switching (Q$_{2Pr}$) charges (P or N = Q$_{leak}$ + Q$_{2Pr}$), while the second pulse response ideally consists of only (U or D = Q$_{leak}$). Thus, subtracting the second pulse response from the first (Q$_{leak}$ + Q$_{2Pr}$ - Q$_{leak}$) theoretically yields Q$_{2Pr}$. However, the conventional 2P$_r$ derived from these methods is not equivalent to partially switched intrinsic 2P$_r$. When the applied voltage is below the coercive voltage, not all polarization domains fully switch during the initial pulse (P or N = Q$_{leak}$ + Q$_{2Pr\_partial}$; Q$_{2Pr\_partial}$ = intrinsic 2P$_r$). This incomplete

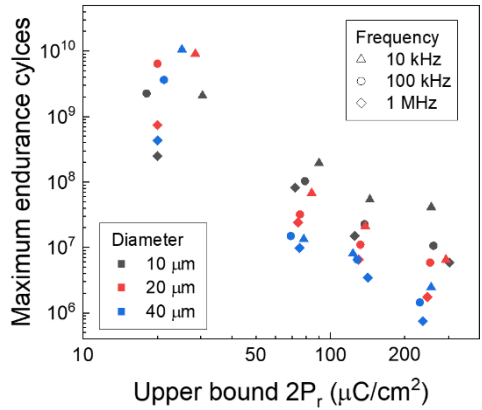

**Supplementary Information Figure S2-2 | The relationship between endurance and upper-bound 2P$_r$ before leakage subtraction.**

switching leaves residual polarization in the material. When the subsequent pulse is applied, this residual polarization contributes to the charge response (U or D = Q$_{leak}$ + Q$_{2Pr\_residual}$). Consequently, the N-D/A calculation underestimates the intrinsic partially switched polarization, resulting in (Q$_{leak}$ + Q$_{2Pr\_partial}$) - (Q$_{leak}$ + Q$_{2Pr\_residual}$) = (Q$_{2Pr\_partial}$ - Q$_{2Pr\_residual}$) instead of the true intrinsic 2P$_r$. Unfortunately, no precise methodology exists to obtain partially switched intrinsic 2P$_r$. Therefore, we adopted an alternative approach by employing P/area (N/area) values instead of P-U/area (N-D/area) in **Figures 3c and 3d**. While not perfect, this method was chosen to resolve the limitations of existing techniques and to provide the most reliable estimation possible under the given constraints.

These parasitic effects become increasingly significant at high applied voltages but are largely negligible when the voltage remains low. As demonstrated in **Figure 3c**, where the preset 2P$_r$ is 10 μC/cm², these interfering factors have a negligible influence due to the low amplitude of the applied voltage. Consequently, the intrinsic 2P$_r$ value is expected to be between 10 and 34.25 μC/cm². Since the effects of leakage and other parasitic elements are insignificant under such conditions, the upper bound of 34.25 μC/cm² serves as a reliable approximation for estimating the partially switched intrinsic 2P$_r$.

Expanding upon this approach, we further examined the conditions necessary for achieving fully switched intrinsic 2P$_r$. Complete polarization switching requires the application of a sufficiently high voltage, which inherently introduces some degree of overestimation. However, before the onset of leakage or material degradation, the capacitor can be driven at relatively low voltages where such overestimation is significantly reduced. Under these conditions, N values remain close to the full 2P$_r$ value of 200 μC/cm². This suggests that our AlScN capacitor exhibits minimal leakage currents and a low density of defects, contributing to reliable electrical performance. The limited influence of parasitic effects supports the validity of the estimation approach used in this study.

**Figure S2-1** provides an analysis of the upper-bound $2P_r$ before and after leakage, utilizing the method outlined in **Figures 3c** and **3d** under various conditions. As frequency increases, the difference between upper-bound $2P_r$ and conventional $2P_r$ diminishes. This is because ferroelectric switching occurs at an inherently high speed, allowing a rapid response at high frequencies. In contrast, leakage effects contribute minimally to overestimation at high frequencies due to the extremely short pulse width time. When the preset $2P_r$ is low, the difference becomes even smaller, which is attributed to the reduced contribution of leakage-induced overestimation at lower polarization states due to relatively low applied voltage. This trend is consistently observed across different diameters.

**Figure S2-2** shows the relationship between endurance and upper-bound $2P_r$ before leakage.

# Supplementary Information S3

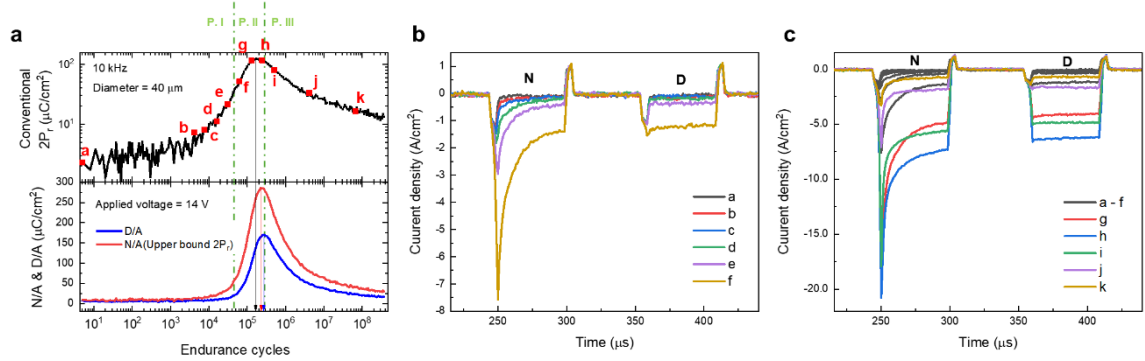

**Supplementary Information Figure S3-1 | The endurance test results under a fixed applied voltage of 14 V, the evolution of polarization and current response across different endurance cycles. (a)** The top panel shows the evolution of conventional $2P_r$ as a function of endurance cycles, with key transition points marked in green. The bottom panel compares the N/A (upper bound $2P_r$) and D/A current densities. **(b)** and **(c)** exhibit the N and D current responses as a function of time in the partial wake up followed by leakage and degradation phase respectively at different endurance cycle moments.

**Supplementary Information S3** provides a clear analysis of how voltage affects ferroelectric switching, dividing it into distinct phases. **Figure S3-1a** shows a case where a consistent 14 V is applied. The conventional $2P_r$ is shown on a logarithmic y-axis, while N/A (upper bound $2P_r$) and D/A are on a linear scale.

The N peak is mainly attributed to ferroelectric polarization switching. However, since the applied voltage is below the $V_C$, the switching does not fully reflect the inherent total spontaneous polarization. Additionally, because the material has not fully woken up, the measured polarization does not intrinsically represent the maximum possible switching. As a result, some residual ferroelectric response appears in the D pulse. The current level is also affected by various factors, but the most dominant factors are the charge accumulation due to the paraelectric properties and the leakage current passing through the insulator. Overall, the N current response mainly includes ferroelectric switching and leakage, while the D current response contains a smaller portion of ferroelectric switching along with leakage.

Through the changes in N/A and D/A, three phases can be distinguished. Phase (P.) P.I, P. II, and P. III, as separated by the green dashed line in **Figure S3-1a**. P. I shows little change in both N/A and D/A, P. II exhibits an exponential increase in N/A and D/A, and P. III marks the beginning of degradation, starting from peak point of D/A. As shown in the traces labeled a-e in **Figure S3-1b**, P. I is characterized by a steady increase only in the peak of the N current response, indicating the wake-up process of AlScN. Since the applied voltage of 14 V is much lower than the $V_C$ of 17 V, the wake-up progresses very slowly, this condition is referred to as partial wake-up. This term describes the gradual activation of ferroelectric switching, where polarization begins to change but has not yet to reach its full potential. The graphs labeled a-e in **Figure S3-1b** also shows that the D pulse remains mostly unchanged, suggesting that leakage does not increase significantly and that the material is still in a partial wake-up stage.

A transition occurs from P. I to P. II, particularly in the traces labeled e to f in **Figure S3-1b**, where an increase in the D current response indicates the beginning of leakage. In the graphs labeled f-h of **Figure S3-1c**, not only partial wake-up but also leakage continues to grow during P. II. The transition from P. II to P. III seen in the traces h-k in **Figure S3-1c** indicates the degradation phase, where both the N and D peak current responses start to decrease due to the material's fatigue phase.

In addition, as seen in **Figure S3-1a**, the peaks of conventional $2P_r$ appear first, followed by the peak of N/A and then D/A. This is because the rates of partial wake-up and leakage increase at different speeds. Partial wake-up happens more quickly and grows faster than leakage during P. I. Whereas in P. II, leakage starts increasing more quickly, indicating a shift in the dominant process. Understanding this interaction between wake-up and leakage provides important insights into the long-term stability and reliability of ferroelectric materials.

**Figure S3-2** shows the case where the applied voltage is fixed at 16 V. Since this voltage is close to $V_C$, a relatively faster wake-up process was observed. Similar to **Figure S3-1**, the region where conventional $2P_r$ increases is also observed in P. I. In **Figure S3-2b**, the traces labeled a-e show no noticeable increase in the D current response, indicating that leakage remains suppressed while partial wake-up is progresses. During P. I, unlike in **Figure S3-1**, the conventional $2P_r$ approaches nearly 200 μC/cm². This value represents full spontaneous polarization while leakage remains minimal. This suggests a mitigated wake-up process, providing insights into optimizing wake-up behavior. Furthermore, unlike in **Figure S3-1**, in P. II, the conventional $2P_r$ exhibits a minimal increase and remains stable. In this phase, N/A and D/A increase at a very slow rate. As observed in the traces labeled f-i in **Figure S3-2c**, the rates of increase for D and N show nearly the same trend, which suggests the system is in a stable phase. Following this stable phase, the system turns into P. III, characterized by an exponential increase of D/A. Eventually, after reaching the peak of D/A, both the N and D peak current responses start to decrease, as shown in **Figure S3-2c** and **S3-2d**.

In **Figure S3-3**, a fixed voltage of 18 V, which is higher than $V_C$, is applied. In this case, no partial wake-up process is observed. As a result, the stable phase that appeared in P. II of **Figure S3-2** directly

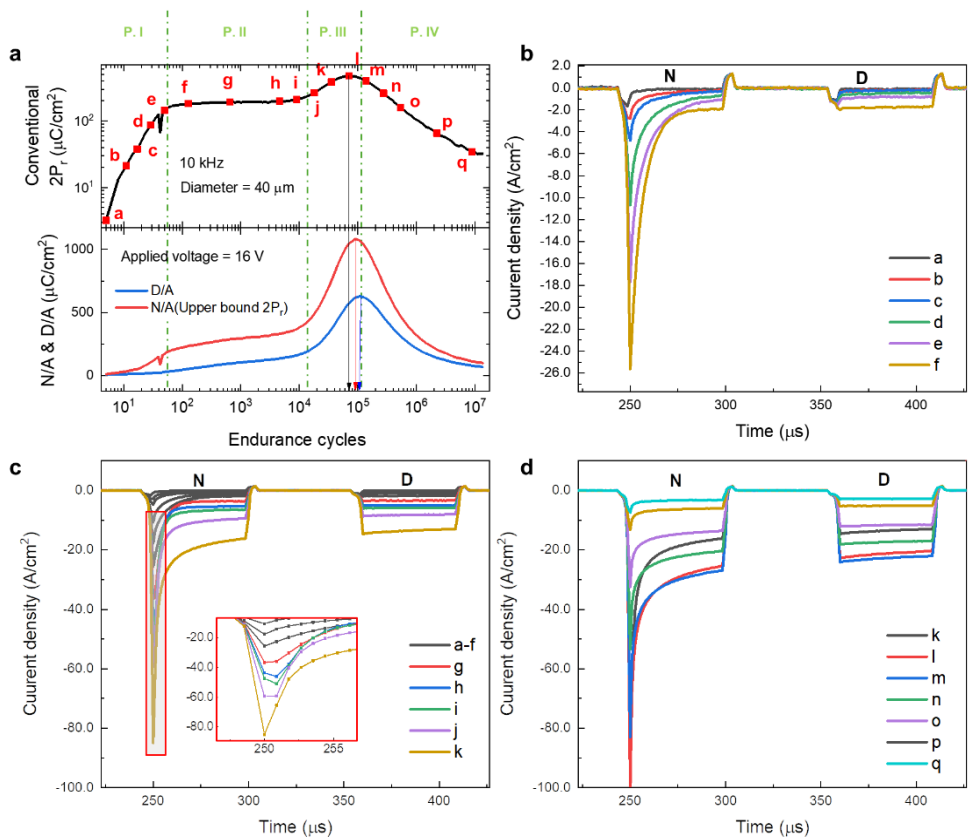

**Supplementary Information Figure S3-2 | The endurance test results under a fixed applied voltage of 16 V, the evolution of polarization and current response across different endurance cycles. (a)** The top panel shows the evolution of conventional $2P_r$ as a function of endurance cycles, with key transition points marked in green. The bottom panel compares the N/A (upper bound $2P_r$) and D/A current densities. **(b-d)** exhibit the N and D current responses as a function of time in the partial wake up, stable and degradation phase respectively at different endurance cycle moments. The inset in **(c)** is a magnified image at the peak of the N current density response.

emerges in P. I of **Figure S3-3a**. As shown in **Figure S3-3b**, traces a–e show a stable state where both the N and D peak current responses continuously increase at the same rate. This stable state is followed by P. II and P. III, where the responses increase exponentially and then decrease.

When a low voltage is applied, the system does not reach its maximum potential value of 200 μC/cm². Instead, it enters a phase where leakage increases without the system stabilizing. Conversely, applying a high voltage causes the system to transition directly into a stable state. However, as seen in **Figure S3-3a**, the D/A peak is approximately ten times higher than that of **Figure S3-1a**. This indicates a significant increase in leakage current due to the high applied voltage. This elevated leakage induces current stress, resulting in a reduction of the achievable endurance cycles. Therefore, applying an

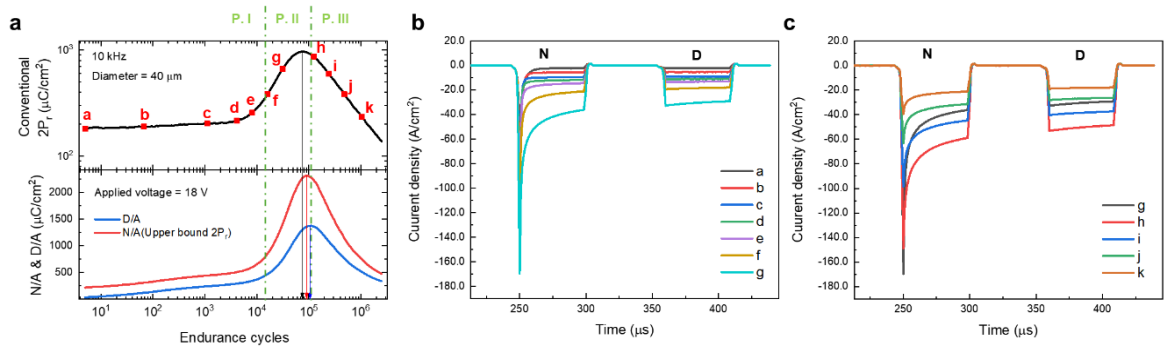

**Supplementary Information Figure S3-3 | The endurance test results under a fixed applied voltage of 18 V, the evolution of polarization and current response across different endurance cycles. (a)** The top panel shows the evolution of conventional $2P_r$ as a function of endurance cycles, with key transition points marked in green. The bottom panel compares the N/A (upper bound $2P_r$) and D/A current densities. **(b-c)** exhibit the N and D current responses as a function of time in the stable and degradation phase respectively at different endurance cycle moments.

optimal voltage serves as an effective strategy to suppress stress and enhance the endurance cycling.

**Figure S3-4** presents the phase transitions occurring due to voltage adjustments, particularly in relation to endurance cycles. In **Figure S3-4a**, a reference line is drawn at $2P_r = 50$ μC/cm². The experimental data points where $2P_r$ intersects this reference line are labeled a-g. Marks a-c correspond to applied voltages of 16 V, 17 V, and 18 V, respectively, indicating the progression of partial wake-up. Marks d-g represent the degradation phase associated with fatigue, where material performance declines.

The markers in **Figure S3-4b** correspond directly to those in **Figure S3-4a**, representing the same conventional $2P_r$ and applied voltage conditions at different endurance cycles. This consistent labeling highlights the correlation between the two figures. The voltage initially increases gradually to bring the conventional $2P_r$ close to the preset $2P_r$. Subsequently, as seen in marks a-c in **Figure S3-4b**, the applied voltage decreases due to partial wake-up, leading to a reduction in operational voltage. This observation aligns with **Figure S3-4a**, where the corresponding region confirms the occurrence of partial wake-up. In contrast, marks d-g correspond to the fatigue phase, where degradation becomes more pronounced. This is evident from the decrease in conventional $2P_r$ in **Figure S3-4a**, while in **Figure S3-4b**, the adjusted applied voltage increases as degradation progresses as the process maintains conventional $2P_r$ close to the preset $2P_r$. These trends indicate that a decreasing voltage corresponds to partial wake-up, whereas an increasing voltage signifies material degradation.

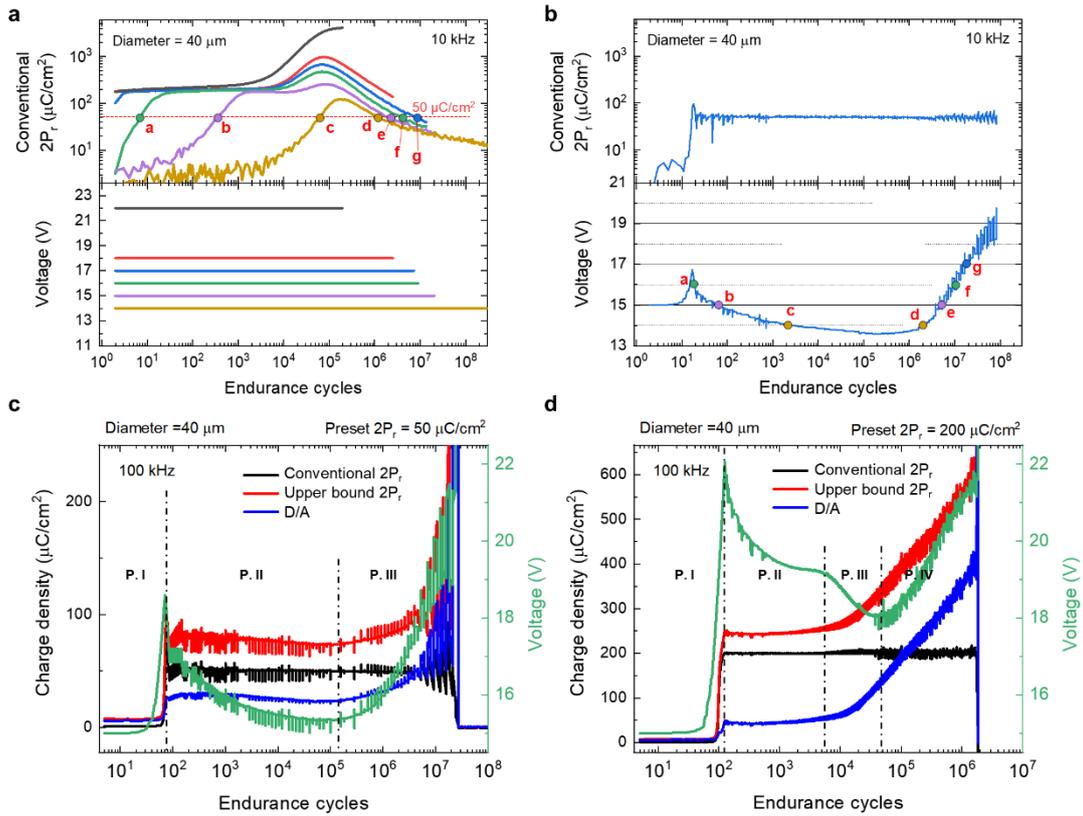

**Supplementary Information Figure S3-4 | Endurance test results related to phase transition for a preset 2P$_r$ of 50 μC/cm².** **(a)** exhibits constant applied voltage pulses with a reference line and **(b)** adjusted applied voltage pulses to maintain a conventional 2P$_r$ of 50 μC/cm². **(c)** and **(d)** show endurance tests under different preset 2P$_r$ values.

A notable observation is the extended interval between marks c and d. By adjusting the voltage, the preset 2P$_r$ level is attained more quickly. As a result, marks a-c in **Figure S3-4b** appear earlier in the endurance cycle compared to their counterparts in **Figure S3-4a**, while marks d-g appear later. Notably, the endurance cycle values for d-g in **Figure S3-4b** are nearly twice those in **Figure S3-4a**, demonstrating an extended device lifespan with optimized voltage adjustment.

**Figure S3-4c** presents the results obtained at 100 kHz, providing a detailed analysis of voltage-dependent phase transitions. Not only conventional 2P$_r$ but also the variations of N/A and D/A are plotted together. During P. I of **Figure S3-4c**, the algorithm rapidly adjusts the applied voltage to achieve the appropriate initial conditions. Subsequently, in P. II, a continuous decrease in applied voltage is observed, which is attributed to partial wake-up effects. During this phase, both N/A and D/A show a slight decreasing trend, as leakage is suppressed due to the reduction in applied voltage. This process is important in delaying ferroelectric breakdown by mitigating electrical stress from the reduced adjusted

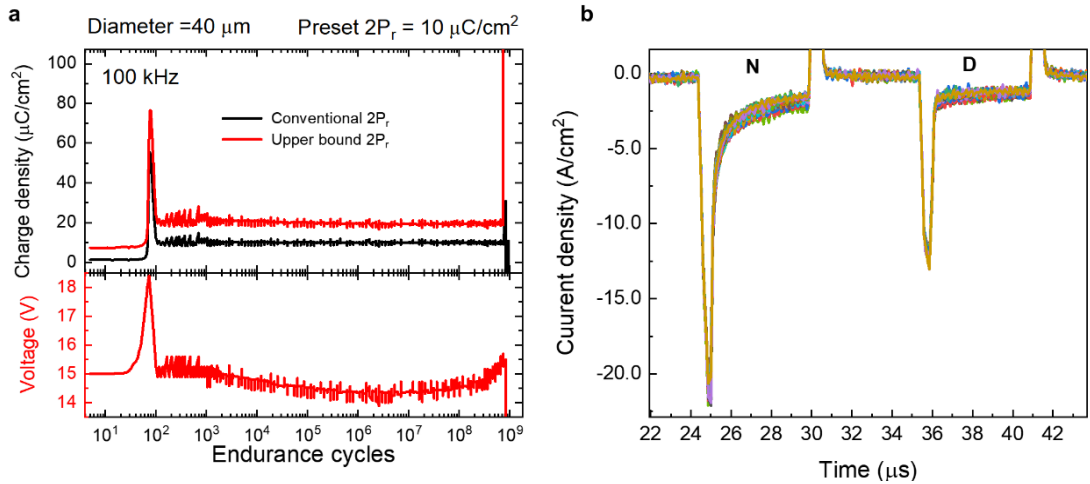

**Supplementary Information Figure S3-5 | Endurance test results and N&D current responses with 40 m diameter under 100 kHz to adjust applied voltage to maintain the conventional 2P$_r$ of 10 μC/cm².**

applied voltage and minimizing current stress resulting from decreased leakage. In P. III, as degradation progresses, leakage increases, leading to a simultaneous rise in both N/A and D/A. Consequently, the applied voltage is also adjusted upward to maintain stable conventional 2P$_r$ switching.

**Figure S3-4d** illustrates the voltage adjustment required to maintain full switching (~200 μC/cm²). In this case, unlike **Figure S3-4c**, four distinct phases can be identified. In P. I of **Figure S3-4d**, the process of finding the appropriate voltage proceeds similarly to **Figure S3-4c**. Following this, in P. II, the applied voltage continues to decrease. During this phase, not only does partial wake-up occur, but the stable phase also continues simultaneously. In P. III, a sharp decrease in applied voltage occurs as N/A and D/A increase, like P. II in **Figure S3-3a**, indicating a phase where leakage rises rapidly. This results in a voltage drop as the system attempts to maintain a constant conventional 2P$_r$. In P. IV, the applied voltage starts to rise, indicating the onset of the fatigue phase caused by ferroelectric degradation.

**Figure S3-5a** shows a diameter of 40 μm with a frequency of 100 kHz and a preset 2P$_r$ of 10 μC/cm². In this case, there is little change in the conventional 2P$_r$ and upper bound 2P$_r$ response throughout the entire endurance test. This indicates that severe leakage does not occur. As a result, additional current stress is minimized. Since only the minimum necessary voltage is applied through the voltage adjustment process, voltage stress is also minimized. Therefore, these minimized stress leads to an increase in endurance. **Figure S3-5b** shows the N and D current density responses from 136 cycles to 5.85E+08 cycles. It confirms that there is no significant change in this response during the entire endurance test. This means that a very stable response is maintained throughout the entire test.

# References

1. S. Fichtner, N. Wolff, F. Lofink, L. Kienle, B. Wagner, AlScN: A III-V semiconductor based ferroelectric., *J. Appl. Phys.*, **125** (11): 114103, (2019)

2. J. X. Zheng, M. M. A. Fiagbenu, G. Esteves, P. Musavigharavi, A. Gunda, D. Jariwala, E. A. Stach, R. H. Olsson, Ferroelectric behavior of sputter deposited $Al_{0.72}Sc_{0.28}N$ approaching 5 nm thickness., *Appl. Phys. Lett.*, **122** (22): 222901, (2023)

3. W. Zhu, J. Hayden, F. He, JI Yang, P. Tipsawat, M. D. Hossain, JP Maria, S. Trolier-McKinstry, Strongly temperature dependent ferroelectric switching in AlN, $Al_{1-x}Sc_xN$, and $Al_{1-x}B_xN$ thin films., *Appl. Phys. Lett.*, **119** (6): 062901., (2021)

4. V. Gund, B. Davaji, H. Lee, J. Casamento, H. G. Xing, D. Jena, A. Lal, Towards Realizing the Low-Coercive Field Operation of Sputtered Ferroelectric $Sc_xAl_{1-x}N$, *2021 21st International Conference on Solid-State Sensors, Actuators and Microsystems (Transducers), Orlando, FL, USA,* 1064-1067, (2021)

5. D. Wang, P. Wang, B. Wang, Z. Mi, Fully epitaxial ferroelectric ScGaN grown on GaN by molecular beam epitaxy., *Appl. Phys. Lett.*, **119**, 111902, (2021)

6. E. Tokumitsu, N. Tanisake, H. Ishiwara, Partial Switching Kinetics of Ferroelectric $PbZr_xTi_{1-x}O_3$ Thin Films Prepared by Sol-Gel Technique, *Jpn. J. Appl. Phys.*, **33,** 5201, (1994)

7. C. Alessandri, P. Pandey, A. Abusleme, A. Seabaugh, Switching Dynamics of Ferroelectric Zr-Doped $HfO_2$, *IEEE Electron Device Letters*, **39**, 1780-1783, (2018)

8. S. Oh, H. Hwang, I. K. Yoo, Ferroelectric materials for neuromorphic computing, *APL Mater.*, **7**, 091109., (2019)

9. R. Guido, H. Lu, P. D. Lomenzo, T. Mikolajick, A. Gruverman, U. Schroeder, Kinetics of N- to M-Polar Switching in Ferroelectric $Al_{1-x}Sc_xN$ Capacitors., *Adv. Sci.*, **11**, 2308797., (2024)

10. K. Yazawa, J. Hayden, JP Maria, W. Zhu, S. Trolier-McKinstry, A. Zakutayev, G. L. Brennecka, *Mater. Horiz.*, **10**, 2936-2944, (2023)